\documentclass[twocolumn,english,aps,pra,showpacs,amsfonts,amsmath,amssymb,superscriptaddress,footinbib,floatfix,longbibliography]{revtex4-1}
\pdfoutput=1
\usepackage[T1]{fontenc}
\usepackage[latin9]{inputenc}
\usepackage{mathtools}
\usepackage{amsmath}
\usepackage{amssymb}
\usepackage{graphicx}
\usepackage{lipsum}
\usepackage{siunitx}
\usepackage{braket}
\makeatletter
\usepackage{ulem}
\usepackage{color}
\usepackage{filemod}

\newcommand{\hidetxt}[1]{}

\usepackage{babel}

\makeatother

\usepackage{babel}

\usepackage{hyperref}

\begin{document}

\title{Nonlinear absorption in interacting Rydberg electromagnetically-induced-transparency spectra on two-photon resonance}

\author{Annika Tebben}
\affiliation{Physikalisches Institut, Universit\"at Heidelberg, Im Neuenheimer Feld 226, 69120 Heidelberg, Germany}

\author{Cl\'{e}ment Hainaut}
\affiliation{Physikalisches Institut, Universit\"at Heidelberg, Im Neuenheimer Feld 226, 69120 Heidelberg, Germany}

\author{Andre Salzinger}
\affiliation{Physikalisches Institut, Universit\"at Heidelberg, Im Neuenheimer Feld 226, 69120 Heidelberg, Germany}

\author{Sebastian Geier}
\affiliation{Physikalisches Institut, Universit\"at Heidelberg, Im Neuenheimer Feld 226, 69120 Heidelberg, Germany}

\author{Titus Franz}
\affiliation{Physikalisches Institut, Universit\"at Heidelberg, Im Neuenheimer Feld 226, 69120 Heidelberg, Germany}

\author{Thomas Pohl}
\affiliation{Center for Complex Quantum Systems, Department of Physics and Astronomy, Aarhus University, DK-8000 Aarhus C, Denmark}

\author{Martin G\"arttner}
\affiliation{Kirchhoff-Institut f\"{u}r Physik, Universit\"{a}t Heidelberg, Im Neuenheimer Feld 227, 69120 Heidelberg, Germany}
\affiliation{Physikalisches Institut, Universit\"at Heidelberg, Im Neuenheimer Feld 226, 69120 Heidelberg, Germany}
\affiliation{Institut f\"ur Theoretische Physik, Ruprecht-Karls-Universit\"at Heidelberg, Philosophenweg 16, 69120 Heidelberg, Germany}
 
\author{Gerhard Z\"urn}
\affiliation{Physikalisches Institut, Universit\"at Heidelberg, Im Neuenheimer Feld 226, 69120 Heidelberg, Germany} 

\author{Matthias Weidem\"uller}
\affiliation{Physikalisches Institut, Universit\"at Heidelberg, Im Neuenheimer Feld 226, 69120 Heidelberg, Germany}

\begin{abstract}
We experimentally investigate the nonlinear transmission spectrum of coherent light fields propagating through a Rydberg EIT medium with strong atomic interactions. In contrast to previous investigations, which have largely focused on resonant control fields, we explore here the full two-dimensional spectral response of the Rydberg gas. Our measurements confirm previously observed spectral features for a vanishing control-field detuning, but also reveal significant differences on two-photon resonance. In particular, we find qualitative deficiencies of mean-field models and rate-equation simulations as well as a third-order nonlinear susceptibility  that accounts for pair-wise interaction effects at low probe-field intensities in describing the nonlinear probe-field response under EIT conditions. Our results suggest that a more complete understanding of Rydberg-EIT and emerging photon interactions requires to go beyond existing simplified models as well as few-photon theories. 
\end{abstract}

\date{\today}

\maketitle
 
\section{Introduction}
\label{sec:intro}

Rydberg electromagnetically induced transparency (EIT) is nowadays a widespread and reliable tool for the creation of strong optical nonlinearities based on Rydberg blockade induced dissipation \cite{FirstenbergAdamsHofferberth:review:JPhysB2016, Pohl:review:16}. Ultimately reaching the regime of quantum many-body nonlinear optics, which requires strong interactions between a large number of photons, would allow one to study fascinating strongly correlated states of light such as photon crystals and quantum fluids of light \cite{Firstenberg:AttractivePhotons:Nature2013,Otterbach:WignerCrystal:PRL13,Bienias:ScatteringResonances:PRA2014,Moos:ManyBodyPolaritons:PRA15,Carusotto:QuantumFluid:RMP13}.

On one hand, increasing the interaction strength per photon is one possible route towards this goal \cite{Chang:QuantumOptics:NatPhot14}. The regime of quantum nonlinear optics, where the optical depth per blockade radius is much larger than one \cite{Pohl:review:16}, has successfully been reached experimentally \cite{Dudin:firstQuantumRydEIT:Science12} and the generation of dissipative \cite{Peyronel:DissipativeQuantum:Nature2012}, attractive \cite{Firstenberg:AttractivePhotons:Nature2013,Liang:ThreePhoton:Science2018,Stiesdal:ThreePhotonCorr:PRL18}, repulsive \cite{Cantu:RepulsivePhotons:NatPhys20}, and spin-exchange-like \cite{Thompson:SymmetryProtectedCollisions:Nature2017} interactions has been demonstrated. Moreover, applications such as single-photon transistors \cite{Tiarks:PhotonTransistor:PRL2014,Gorniaczyk:PhotonTransistor:PRL2014} and gates \cite{Tiarks:QuantumGate:NatPhys19} became experimentally feasible. Beyond this, an experimental study of quantum nonlinear effects with an increased number of photons has been reported \cite{Bienias:HighInput:PRR20}. 

On the other hand, understanding high intensity Rydberg EIT, where the number of photons is large, is the other precursor of quantum many-body nonlinear optics with photons. This regime, where light fields can be treated classically, is where the experimental investigation of Rydberg
EIT was initiated by the demonstration of strong nonlinearities in a Rydberg gas under EIT conditions \cite{Pritchard:firstEITmot:PRL10}. Thereafter, the role of interactions in Rydberg EIT was in detail \cite{Schempp:CPT:PRL10,Sevincli:Adams:CPTEIT:JoPB11,Han:ShiftAndDephasing:PRA2016,DeSalvo:MeanField:PRA2016} with measurements where either the probe or the control field resonantly couples its atomic transition and various theoretical models have been put forward to describe the experimental observations. While the developed theory has been successful in reproducing the general effects of Rydberg-state interactions on EIT in such systems, important spectral details such as the appearance and origins of nonlinear shifts and asymmetries are still under debate \cite{Sevincli:Adams:CPTEIT:JoPB11,Han:ShiftAndDephasing:PRA2016}.

Here, we address this question by studying the spectral properties of EIT transmission in a cold atomic gas with strong Rydberg-state interactions. We choose to work in the Autler-Townes regime \cite{DeSalvo:MeanField:PRA2016} with rubidium atoms in the $48$S Rydberg state and broadly scan both control- and probe-field frequencies. In particular, we measure the nonlinear absorption spectrum of the probe field on two-photon resonance, i.e., by simultaneously adjusting the control-field frequency to maintain EIT conditions. As this implies a vanishing linear absorption for all probe-field frequencies, such measurements directly probe pure interaction effects. In turn, the measured spectra reveal qualitative discrepancies with existing theories that were previously missing in experiments with resonant control fields \cite{Sevincli:Adams:CPTEIT:JoPB11,Han:ShiftAndDephasing:PRA2016,DeSalvo:MeanField:PRA2016}. Existing mean-field approximations and semiclassical rate-equation simulations, as well as a low-intensity expansion \cite{Sevincli:adibatic:PRL11,Tebben:ResonantEnhancement:PRA19}, which accounts for resonances with control-laser dressed entangled pair states \cite{Gaerttner:Resonance:PRL14,Gaul:ResDressing:PRL16,Helmrich:ResDressing:PRL16,Tebben:ResonantEnhancement:PRA19}, fail qualitatively to reproduce the observed broad absorption features away from single-photon resonance suggesting the need to go beyond existing simplified models of Rydberg EIT.

This article is organized as follows. Following a brief description of the experimental setup in Sec. \ref{sec:experiment}, we present the main results of the nonlinear transmission measurements in Sec. \ref{sec:scan2photRes-meas}. In Sec. \ref{sec:theory} we compare the observations with the prediction of a mean-field approximation \cite{Weatherill_2008,DeSalvo:MeanField:PRA2016,Han:ShiftAndDephasing:PRA2016},  Monte Carlo rate-equation simulations \cite{Ates:ExcitationDynamicsRydberg:PRA07,Ates:EITuniversality:PRA2011,Heeg:Evers:HybridModel:PRA12,Gaerttner:REwithPropagation:PRA13} and a theory based on the third-order susceptibility that accounts for pair-wise atomic correlations \cite{Sevincli:adibatic:PRL11,Tebben:ResonantEnhancement:PRA19}. In Sec.~\ref{sec:theory_comparison} we
compare the theoretical models with experimental measurements. We summarize in Sec.~\ref{sec_concl}.

\begin{figure}[t!]
	\includegraphics[width=1\linewidth]{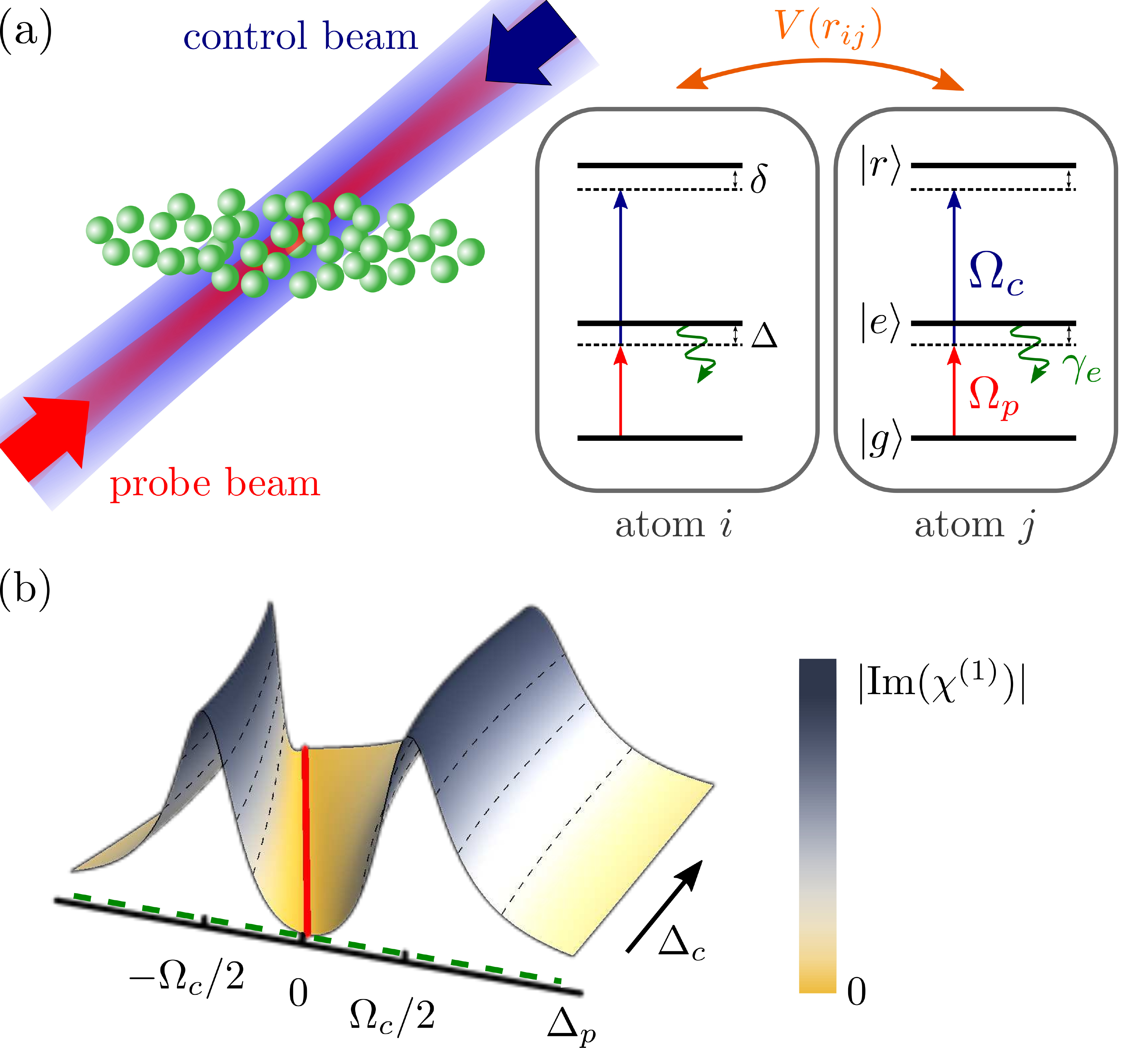}
	\caption{(a) Realization of a ladder-type Rydberg EIT system with counter-propagating probe (red) and control (blue) beams. Indicated detunings in the atomic level scheme are the single-photon detuning $\Delta=\Delta_p$ and two-photon detuning $\delta=\Delta_p+\Delta_c$, which are given by the laser detunings $\Delta_{p,c}$ of the probe and control beams, respectively. Rubidium Rydberg atoms interact via an interaction $V(r_{ij})$, which depends on the distance $r_{ij}$ between two atoms $i$ and $j$. (b) Imaginary part of the linear optical response $|\text{Im}(\chi^{(1)})|$ as a function of the laser detunings. Transmission spectra are measured, where the control beam detuning $\Delta_c=0$ (green dashed line), or where only the single-photon detuning $\Delta$ is changed while staying on two-photon resonance ($\delta=0$, red line). Here, the optical response vanishes in the non-interacting regime.
	}
	\label{fig:level-scheme_chi1} 
\end{figure}

\section{The Rydberg EIT setup}
\label{sec:experiment}
Our Rydberg EIT medium consists of a cigar-shaped $^{87}$Rb atom cloud with $1/e^2$-waists of about $40\times40\times\SI{300}{\mu m^3}$ held within an optical dipole trap, with a maximal peak density of $\SI{2e11}{cm^{-3}}$ and with a temperature of approximately $\SI{100}{\mu K}$. As detailed in App. \ref{app-sec:exp_details} we ensure an accurate preparation of the ground state $\ket{g}= \ket{5\text{S}_{1/2},F=2,m_F=2}$, which together with the short-lived intermediate state $\ket{e}=\ket{5\text{P}_{3/2},F=3,m_F=3}$ with decay rate $\gamma_e/2\pi=6.067$ MHz and a metastable Rydberg state $\ket{r}=\ket{48\text{S}_{1/2}, m_j=1/2}$ forms the three-level ladder EIT system, as depicted in Fig.~\ref{fig:level-scheme_chi1}(a). Rydberg atoms interact via isotropic van der Waals interactions $V(r_{ij})=C_6/r_{ij}^6$, where $C_6$ is the van der Waals coefficient and $r_{ij}$ the distance between two atoms $i$ and $j$. Counter propagating probe ($\SI{780}{nm}$) and control ($\SI{480}{nm}$) beams with Rabi frequencies $\Omega_{p}$ and $\Omega_c$, couple the ground to intermediate and intermediate to Rydberg-state transitions, respectively. The optical depth per blockade radius $\text{OD}_b$ is much smaller than $1$ in our system, meaning that the probe field preserves its coherent nature and can be treated classically \cite{Pohl:review:16}.

Earlier investigations of EIT spectra revealed, that the EIT area can act as a dispersive gradient-index lens \cite{Li:EITlensing:PRA15}. This is a consequence of a refractive index gradient, which is induced by the inhomogeneity of a focused control beam within a uniform probe beam. In order to avoid this effect and to ensure negligible dispersion, we invert this geometry and choose a control beam with a waist twice as large as the focused probe beam ($1/e^2$ beam waist: $\SI{15}{\mu m}$).

After releasing the atoms from the dipole trap, probe and control beam pulses are applied in order to create EIT conditions. We experimentally checked, that the used probe power and pulse times of $5$ to $\SI{15}{\mu s}$ are small enough to avoid an avalanche creation of Rydberg ions \cite{RobertVincent:AvalancheIonization:PRL13}. Moreover, in order to adiabatically transfer the atoms into the EIT dark-state \cite{Fleischhauer:EIT:PMP05}, we switch on the control beam about $\SI{2}{\mu s}$ before we turn on the probe beam pulse. This is in combination with a rise time of the probe beam pulse of $\SI{50}{ns}$, given by the acousto-optic modulator rise time, sufficient for an adiabatic preparation of the EIT-dark state.

After the EIT sequence we image the transmitted probe light onto a CCD camera, using a 4f-imaging system with a resolution of approximately $\SI{5}{\mu m}$ (Rayleigh criterion). The pixel size of the CCD camera in the imaging plane is $\SI{2.1}{\mu m}$, which is small compared to the waist of the two laser beams. This allows us to average the signal of up to $4\times4$ pixels, corresponding to a maximal variation of $9\%$ in the probe beam Rabi frequency, in order to obtain a better signal-to-noise ratio for the determination of the probe beam transmission $T$. The latter is defined by the ratio of the measured transmitted light in the presence and absence of the atomic cloud.

\section{Transmission measurements}
\label{sec:scan2photRes-meas}
In order to benchmark our Rydberg EIT system against existing measurements \cite{Pritchard:firstEITmot:PRL10,Sevincli:Adams:CPTEIT:JoPB11,Han:ShiftAndDephasing:PRA2016,DeSalvo:MeanField:PRA2016}, we first record the probe beam transmission as a function of the probe beam detuning $\Delta_p$  while staying on resonance with the control laser ($\Delta_c=0$). This measurement follows the green dashed line depicted in Fig.~\ref{fig:level-scheme_chi1}(b) and results in what is typically called an Autler-Townes spectrum. Afterwards, we present measurements on two-photon resonance by experimentally following the red line in Fig.~\ref{fig:level-scheme_chi1}(b). Here, an alternative dimension of investigating Rydberg EIT is pursued, as the linear response of the medium vanishes in the noninteracting regime.

\subsection{Non-interacting and interacting Autler-Townes spectra}
\begin{figure}[t!]
	\includegraphics[width=0.95\linewidth]{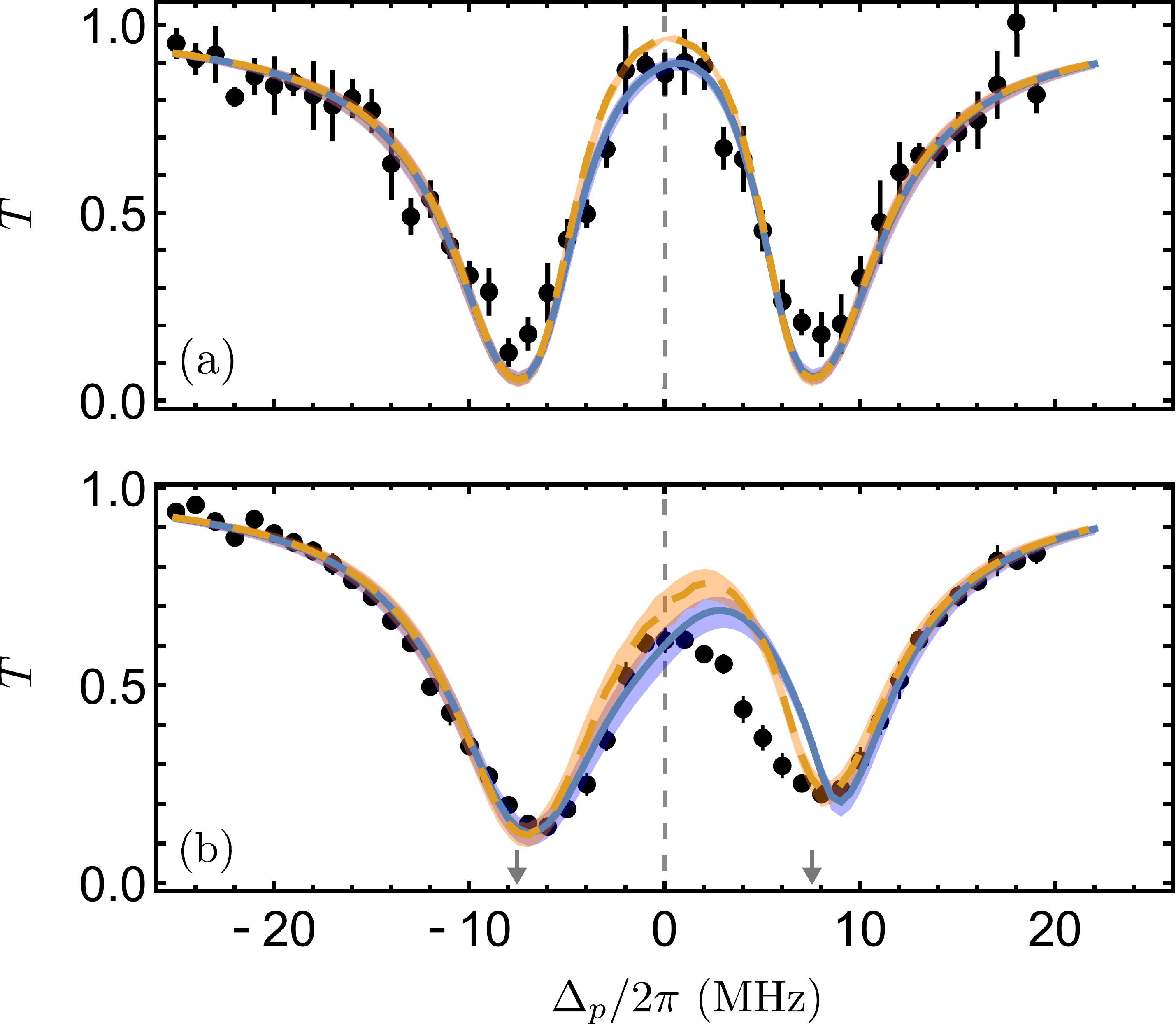}
	\caption{Autler-Townes spectra in the (a) non-interacting and (b) interacting regime. The measured probe beam transmission $T$ against the probe detuning $\Delta_p$ is shown (black points) for (a) $\Omega_p/\Omega_c = 0.03$ and (b) $\Omega_p/\Omega_c=0.12$ for $\Delta_c=0$ and a probe pulse duration of $\SI{15}{\mu s}$. The transmission spectra calculated with a mean-field model (blue solid line) and a MCRE simulation (orange dashed line) are depicted for a fitted peak atomic density of $\SI[separate-uncertainty]{0.16}{\mu m^{-3}}$. Shaded areas take into account the uncertainty in the atomic density. For both theoretical curves $\Omega_c/2\pi=\SI{15}{MHz}$ and $\gamma_{ge}/2\pi= \SI[separate-uncertainty]{1.4}{MHz}$. Gray arrows in (b) indicate the minimum positions of the curve in (a).}
	\label{fig:AutlerTownes} 
\end{figure}
When changing the probe beam detuning $\Delta_p$ with the control beam on resonance ($\Delta_c=0$) in the non-interacting regime, where the probe Rabi frequency is small, we recover the known Autler-Townes spectrum, as shown in Fig.~\ref{fig:AutlerTownes}(a). A transmission of nearly $1$ at zero detuning and a symmetric spectrum supports negligible dephasing $\gamma_{gr}$ on the Rydberg coherence and therefore a largely coherent dynamics of a three-level system.

In the interacting regime at a high probe Rabi frequency, as presented in Fig.~\ref{fig:AutlerTownes}(b), the transmission at zero detuning is reduced. Moreover, we observe a small shift of the left minimum and an asymmetry of the spectrum. Solid and dashed lines in Fig.~\ref{fig:AutlerTownes} are a the result of theoretical models and will be discussed in Sec. \ref{sec:theory}. 

In the first experimental demonstration of nonlinearities in a Rydberg EIT medium, no shift and asymmetry were measured \cite{Pritchard:firstEITmot:PRL10} in the EIT spectrum. This was explicitly attributed to the absence of Rydberg excitations or ions in the system, which could cause a mean-field shift, and was explained as a sole cooperative nonlinearity. Subsequent publications, even of the same group, showed measurements that exhibited both a shift and asymmetry in the EIT as well as the Autler-Townes regime \cite{Han:ShiftAndDephasing:PRA2016,DeSalvo:MeanField:PRA2016,Sevincli:Adams:CPTEIT:JoPB11}. There, the absence of the asymmetry in earlier measurements was discussed as the result of increased absorption due to interaction-induced motion in the time of one experimental cycle \cite{Sevincli:Adams:CPTEIT:JoPB11}. We cannot fully exclude motional dephasing
originating from interaction-induced repulsion of the
atoms in our system. However, we still observe the asymmetry
in contrast to ref.~\cite{Sevincli:Adams:CPTEIT:JoPB11}. This already shows that features of Rydberg EIT spectra, with the control beam on resonance, are not yet fully understood.

\subsection{Measurements on two-photon resonance}
\begin{figure}[t!]
	\includegraphics[width=0.95\linewidth]{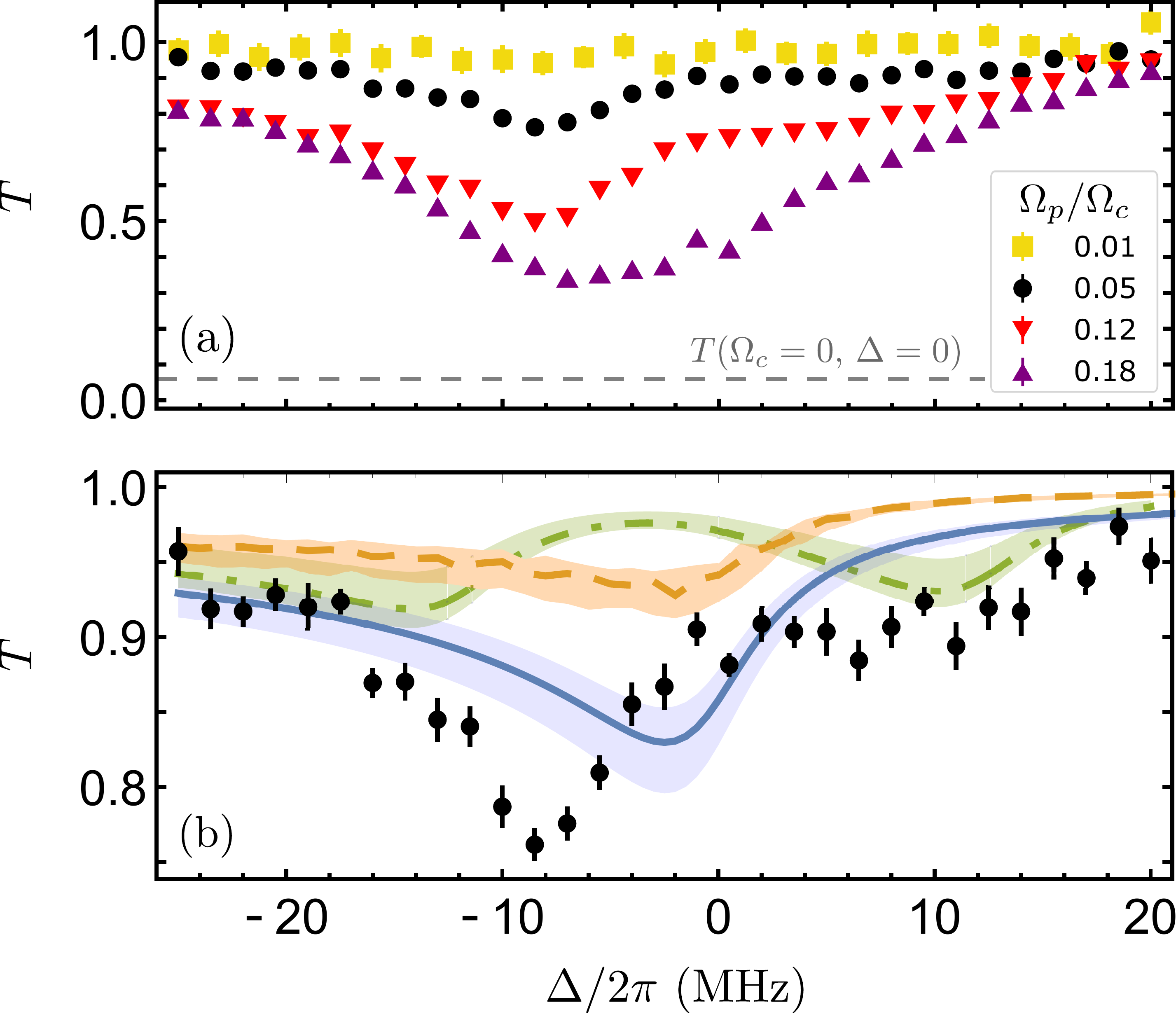}
	\caption{Transmission $T$ on two-photon resonance ($\delta=0$). (a) Measured transmission spectra for different ratios $\Omega_p/\Omega_c$. The gray dashed line indicates the measured transmission in the absence of the control beam at $\Delta=0$. Here, $\Omega_c/2\pi=\SI[separate-uncertainty]{28}{MHz}$ for the yellow curve and $\SI[separate-uncertainty]{24}{MHz}$ for all other curves. The probe pulse duration is $\SI{5}{\mu s}$ for all measurements.
	(b) Comparison of the measured transmission spectrum at $\Omega_p/\Omega_c=0.05$, with the results of the mean-field model (blue solid line), the MCRE simulation (orange dashed line) and the low-intensity theory (green dash-dotted line) for a peak atomic density of $\SI[separate-uncertainty]{0.16}{\mu m^{-3}}$ and with $\gamma_{ge}/2\pi= \SI[separate-uncertainty]{1.4}{MHz}$. Shaded areas take into account the uncertainty in the atomic density.}
	\label{fig:scan2photRes} 
\end{figure}
In order to investigate the effect of interactions on the transmission spectrum in a different approach we perform measurements on two-photon resonance by choosing experimental parameters to follow the red line in Fig.~\ref{fig:level-scheme_chi1}(b), where the linear response of the medium vanishes in the non-interacting regime. Precisely, we change the single-photon detuning $\Delta$ while staying on two-photon resonance ($\delta=0$) and record the probe beam transmission as presented in Fig.~\ref{fig:scan2photRes}(a).

At low $\Omega_p$ the probability to be in the Rydberg state is small, such that in this so-called non-interacting regime nonlinearities due to Rydberg interactions are negligible. Here, the transmission is consistent with unity for all single-photon detunings, as shown by the yellow squares in Fig.~\ref{fig:scan2photRes}(a). This is expected as on two-photon resonance the linear response of the atomic medium vanishes and EIT conditions are fulfilled.

Gradually increasing the probe Rabi frequency increases the Rydberg state fraction, such that interaction effects influence the dynamics. In this interacting regime, already for a ratio of $\Omega_p/\Omega_c\approx 0.05$, depicted by the black circles in Fig.~\ref{fig:scan2photRes}(a), a dip in the transmission to about $0.76$ appears on the negative detuning side. However, this dip is absent on the positive detuning side. For the experimental parameters of this measurement we can exclude that this absorption feature results from the influence of stationary Rydberg excitations or ions in the medium. In order to support this statement, we present a measurement of Rydberg excitations after the EIT sequence and an estimation of an upper limit for the absorption resulting from these in App. \ref{app-sec:ion_measurement}.

Increasing the probe Rabi frequency further (red down triangles) increases the strength of the absorption dip, but does not change its position. Moreover, the feature is getting broader, but remains clearly visible until $\Omega_p/\Omega_c\approx0.12$. For the highest ratio $\Omega_p/\Omega_c\approx 0.18$ of the two Rabi frequencies (purple up triangles) strong absorption continues to persists predominantly on the negative detuning side but is shifted towards the single-photon resonance and is further broadened. Therefore, the observed absorption feature turns out to be very sensitive to $\Omega_p$, which is a characteristic of a nonlinear phenomenon.

\section{Theoretical models}
\label{sec:theory}

Having observed a small shift and an asymmetry in the interacting Autler-Townes spectrum and a strong nonlinear absorption in spectra on two-photon resonance, we now aim for a comparison of our measurements with existing theoretical models. Working in the regime, where the probe field can be treated classically ($\text{OD}_b\ll1$), photon-photon and atom-photon correlations can be neglected \cite{Pohl:review:16}. However, interactions between atoms induce atom-atom correlations, which on one hand enable strong nonlinear effects and therefore make the system potentially useful for quantum optics applications, but on the other hand make a theoretical treatment of the Rydberg EIT medium challenging.
Exactly and numerically only solvable for a few atoms \cite{Pritchard:firstEITmot:PRL10}, methods to truncate these correlations in the resulting many-body master equations \cite{Ates:ExcitationDynamicsRydberg:PRA07,Sevincli:Adams:CPTEIT:JoPB11} need to be applied.

The simplest truncation neglects direct two-body correlations in a mean-field approach and implements the resulting interaction-induced shift and dephasing into the single-body master equation \cite{Han:ShiftAndDephasing:PRA2016,DeSalvo:MeanField:PRA2016,Hsiao:MFnearestNeighbor:20}. The same ansatz of an interaction-induced shift is followed in a rate-equation model \cite{Ates:ExcitationDynamicsRydberg:PRA07,Ates:EITuniversality:PRA2011,Heeg:Evers:HybridModel:PRA12,Gaerttner:REwithPropagation:PRA13}, which however solves the many-body rate equations using a Monte Carlo simulation. Another approach for truncation of the many-body correlations is a low-intensity approximation of the optical Bloch equations, in which two-body interactions can be treated exactly \cite{Sevincli:adibatic:PRL11,Tebben:ResonantEnhancement:PRA19}. In the following we give some relevant details on these three approaches and comment on their range of validity.

\subsection{Mean-field model}
Among other implementations of the mean-field model, we choose to compare our experimental results with an ansatz followed in \cite{Han:ShiftAndDephasing:PRA2016}, as experimental parameters, such as the ratio $\Omega_p/\Omega_c$ and the atomic density, are similar. In that work, the transmission of the probe field follows from the one-dimensional Maxwell Bloch equation, where the optical response of the medium enters in terms of a model susceptibility $\bar\chi=\alpha\chi_B+(1-\alpha)\chi_E$ \cite{Han:ShiftAndDephasing:PRA2016}. This model susceptibility is based on the solution for the susceptibility $\chi_\text{3lvl}$ of the single-atom master equation for a non-interacting three-level system and includes interactions in terms of level shifts. Two different parts $\chi_B$ and $\chi_E$ of the model susceptibility are weighted according to the fraction $\alpha$ of all blockaded atoms excluding Rydberg atoms \cite{Han:ShiftAndDephasing:PRA2016}.

Thereby, $\chi_B$ describes the optical response of blockaded atoms and is given by a spatial integration of $\chi_\text{3lvl}(\Delta_c'=\Delta_c+C_6/r^6)$ over the radius $r$ inside the blockade radius. Here, interactions are induced as a level shift and $\chi_B$ equals the two-level susceptibility for strong interactions. In addition, $\chi_E=\chi_\text{3lvl}(\Delta_c'=\Delta_c-\Delta_R, \gamma_{gr}=\sqrt{\theta_R})$ accounts for interactions of unblockaded atoms with Rydberg excitations at a distance larger than the blockade radius by introducing an average shift $\Delta_R$ and its variance $\theta_R$, where the latter leads to a dephasing $\gamma_{gr}$ of the Rydberg coherence \cite{Han:ShiftAndDephasing:PRA2016}. In this model $\Delta_R$ and $\theta_R$ can only be approximated and are based on a local density approximation and an approximation for the Rydberg excited fraction, that is derived from a semi-analytical model  using superatoms \cite{Gaerttner:SemianalyticalModel:PRA14}.

\subsection{Monte-Carlo rate equation model (MCRE)}
In this approach, the single-atom master equation without interactions is cast into a set of rate equations by adiabatically eliminating the coherences \cite{Ates:ExcitationDynamicsRydberg:PRA07,Ates:EITuniversality:PRA2011,Heeg:Evers:HybridModel:PRA12}. Interactions are included as effective level shifts $\Delta_\text{int}^{(i)}=\sum_{j\neq i}C_6/r_{ij}^6$
for the Rydberg level of the $i$-th atom with distance $r_{ij}$ to atom $j$. Using a Monte-Carlo simulation, the many-body problem is solved by propagating the global ground state to the global steady state. In this Monte Carlo rate equation (MCRE) model the propagation of the probe field can be included by taking into account the local probe Rabi frequency $\Omega_p^{(i)}$, which atom $i$ experiences, for the calculation of the steady state of atom $i$ in each Monte-Carlo step \cite{Gaerttner:REwithPropagation:PRA13}. For this purpose, the probe Rabi frequency is propagated through the cloud of randomly positioned atoms according to the one-dimensional Maxwell-Bloch equation until atom $i$ is reached. Thereby, in each propagation step, the local attenuation experienced by the individual atoms that are passed is subsequently accounted for. As a result, not only global atomic observables, but also the probe beam transmission can be simulated with this approach.

\subsection{Low-intensity theory}\label{sec:chi3}
A more rigorous description of Rydberg EIT can be obtained by expanding the many-body problem of the interacting ensemble in terms of the number of Rydberg excitations per blockade volume \cite{Sevincli:adibatic:PRL11,Pohl:review:16,Gaul:ResDressing:PRL16,Tebben:ResonantEnhancement:PRA19}. Starting with the underlying Heisenberg equations that describe the driven dynamics of the atomic states one obtains a hierarchy of equations for operator products that describe correlations and entanglement induced by the strong Rydberg-state interactions. Assuming that the Rydberg population per blockade radius is small, this hierarchy can be truncated by neglecting three-body contributions, i.e., assuming that the probability to excite three nearby strongly interacting Rydberg atoms is negligibly small. This permits us to find an exact analytic solution for the third-order nonlinear susceptibility that fully accounts for two-body correlations and entanglement on the level of atomic pairs \cite{Tebben:ResonantEnhancement:PRA19}. The transmission of the probe beam is then calculated by applying a local density approximation and assuming spatially constant probe and control beams. 

\subsection{Range of validity}
For a comparison of the range of validity of the theoretical models the strength of the applied fields, the atomic density, and the interaction strength are considered in the following.

The mean-field model includes Rydberg interactions solely as an interaction-induced energy shift based on the assumption that interatomic correlations can be completely neglected. This requires that the mean distance between Rydberg excitations is larger than the blockade radius, which is the case, for example, if $\Omega_p/\Omega_c\ll1$ or for a small interaction strength. A simple mean-field model has been shown to fail to explain observations in coherent population trapping experiments as soon as excitation blockade becomes relevant \cite{Schempp:CPT:PRL10}, which depending on the experimental parameters might already be the case at low atomic densities. The mean-field model considered here agreed well with an interaction induced shift and dephasing observed in interacting EIT transmission spectra for densities up to about $\SI{0.1}{\mu m^{-3}}$ \cite{Han:ShiftAndDephasing:PRA2016}.

The MCRE model also includes Rydberg interactions as effective energy shifts but does not rely on assumptions for calculating the average shift experienced by one atom. Instead, it naturally includes the mean-field shift in a self-consistent manner and calculates the steady-state of the $N$-body density matrix. While still requiring $\Omega_p/\Omega_c\ll1$ or $\Omega_p/\Omega_c\gg1$ for atomic coherences to vanish \cite{Gaerttner:SemianalyticalModel:PRA14},
these two aspects increase its range of validity to a large range of atomic densities and yields correct results also for densities as high as $\SI{0.18}{\mu m^{-3}}$\cite{Ates:EITuniversality:PRA2011,Sevincli:Adams:CPTEIT:JoPB11}. For $\Omega_p/\Omega_c\ll 1$ the rate-equation model was shown to agree well with the result of a master equation calculation of four fully blockaded atoms independently of the driving strength $\Omega_c/\gamma_e$ \cite{Gaerttner:SemianalyticalModel:PRA14}. As a result, the MCRE model was able to explain certain aspects of interacting EIT spectra and the density dependence of nonlinear absorption \cite{Sevincli:Adams:CPTEIT:JoPB11,Gaerttner:REwithPropagation:PRA13}. 
 
The low-intensity theory is based on a perturbative expansion in the probe field and therefore requires $\Omega_p/\Omega_c\ll 1$. Its applicability in terms of atomic densities and interaction strength is combined in the requirement, that the Rydberg population per blockade volume needs to be much smaller than $1$ \cite{Pohl:review:16}. In its form considered here, the low-intensity theory predicted the existence of an enhanced nonlinear optical response for $\Delta\sim\pm\Omega_c/2$ as a consequence of a two-body two-photon resonance in the nonadiabatic regime \cite{Tebben:ResonantEnhancement:PRA19}, but has not been compared to experiments in this regime yet. However, in the regime of large probe beam detunings, where the intermediate state can be adiabatically eliminated, the low-intensity theory was successfully compared to absorption measurements showing the quadratic dependence on the probe Rabi frequency at moderate densities \cite{Sevincli:adibatic:PRL11}.

\section{Comparison between theory and experiment}
\label{sec:theory_comparison}
We implement the three models \cite{Han:ShiftAndDephasing:PRA2016,Gaerttner:REwithPropagation:PRA13,Tebben:ResonantEnhancement:PRA19} with a transversely constant probe beam intensity and a constant control beam intensity in all spatial dimensions. All three models account for the $45°$ angle between the propagation direction of the lasers and the main axis of the atomic cloud, as depicted in Fig.~\ref{fig:level-scheme_chi1}(a), and include the Gaussian density distribution in the propagation direction. In the MCRE model the Gaussian density distribution in the transversal direction is considered, while it is assumed to be constant for the other two models. This approximation is justified by the probe beam waist being small compared to the cloud dimension in the transversal direction.  We checked that in the absence of interactions all three models coincide with each other.

\subsection{Comparison with Autler-Townes measurements}
The Rydberg population per blockade volume of the interacting Autler-Townes measurement is $0.16$ on resonance and increases off-resonance even further such that it cannot be considered much smaller than $1$. Therefore, we omit a comparison of the low-intensity theory with the Autler-Townes measurement in the following. The ratio of the two Rabi frequencies as well as the atomic density are in a regime where a comparison with the other two models is possible.

In order to match the result of the mean-field model and the MCRE simulation to our measured transmission spectra in Fig.~\ref{fig:AutlerTownes}, we use for both models a dephasing $\gamma_{ge}/2\pi= \SI[separate-uncertainty]{1.4(5)}{MHz}$ of the excited state coherence to account for a density dependent dephasing present in the system (see App. \ref{app-sec:2lvl_dephasing} for details). Dephasing due to laser noise was independently determined in a measurement of the two-photon linewidth and found to be  $\gamma_{gr}=\SI[separate-uncertainty]{33(4)}{kHz}$. We determine the peak atomic density with a fit to the Autler-Townes spectrum in the non-interacting regime and find $\SI[separate-uncertainty]{0.16(2)}{\mu m^{-3}}$ with a systematic error of $+4\%$ (see App. \ref{app-sec:2lvl_dephasing}). The uncertainty in the density is included in Fig.~\ref{fig:AutlerTownes} as a shaded area. 

In the noninteracting regime the mean-field model (blue solid line) and the result of the MCRE simulation (orange dashed line) agree well with the measured transmission spectrum in Fig.~\ref{fig:AutlerTownes}(a).  The slight deviations observed can come from a small fluctuation of the control beam power and a possible small misalignment of the counter-propagating beams, which are not included in the theoretical models.

In the interacting regime, shown in Fig.~\ref{fig:AutlerTownes}(b), a reduction of the transmission around $\Delta_p=0$ and an asymmetry in the spectrum are predicted by the two models. However, while both theories predict a shift of the resonance position to $\Delta_p/2\pi\approx\SI{3}{MHz}$, we do not observe this large shift in the experiment. The deviation of the transmission predicted by the two theories around single-photon resonance can be explained by the different implementation of the interaction-induced level shift and its variance in the two models (see App. \ref{app-sec:mean-field_MCRE} for details). Moreover, the observed lower transmission on resonance can be captured by including an effective dephasing rate $\gamma_{gr}$ in the models, which could be explained by Rydberg excitations that might be present in the medium for the experimental parameters of this measurement as discussed in App. \ref{app-sec:ion_measurement}.

While the attenuation of the transmission on resonance is a consequence of Rydberg blockade-induced absorption and an experimentally and theoretically approved feature of interacting Rydberg EIT systems \cite{Pohl:review:16}, the absence or presence of a shift and asymmetry in the spectrum are debated in the literature \cite{Sevincli:Adams:CPTEIT:JoPB11,Han:ShiftAndDephasing:PRA2016}. In theories, that rely on a mean-field shift of the Rydberg level, such as the considered mean-field and MCRE models, the asymmetry and shift are a consequence of an anti-blockade effect. It allows the excitation of Rydberg pair states for a positive probe detuning, thereby reducing absorption and effectively shifting the resonance position \cite{Sevincli:Adams:CPTEIT:JoPB11}. This shift is also observable in the solution of the master equation for a few atoms \cite{Gaerttner:REwithPropagation:PRA13}. As discussed in Sec. \ref{sec:scan2photRes-meas} A, the asymmetry and shift might be reduced in setups where interaction induced atomic motion moves atoms out of the facilitation resonance \cite{Sevincli:Adams:CPTEIT:JoPB11}.

\subsection{Comparison with measurements on two-photon resonance}
For the measurements on two-photon resonance the atomic density is in a regime where all three models should be applicable. Moreover, the Rydberg population per blockade volume is below $0.05$ for the yellow and black curves in Fig.~\ref{fig:scan2photRes}(a) for all detunings, but exceeds this threshold for the other two curves. This means that at least for the yellow and black curves, for which $\Omega_p/\Omega_c\ll1$, the requirements for all three models are fulfilled.

In the non-interacting regime on two-photon resonance ($\delta=0$), the transmission is nearly $1$ for all single-photon detunings $\Delta$, as shown by the yellow squares in Fig.~\ref{fig:scan2photRes}(b). This shows that experimental imperfections, which would lead to single-particle dephasing (e.g. atomic motion, imperfect initial state preparation and remnant DC electric fields), are negligible. In theoretical models a transmission of $1$ is expected, as on two-photon resonance the population in the Rydberg state, and thus interaction-induced shifts, tends to zero for small $\Omega_p$ or small atomic densities. In combination with $\chi_\text{3lvl}(\delta=0)=0$ for negligible single-particle dephasing this results in a vanishing linear response. All three models reproduce this behavior correctly.

For the interacting regime, Fig.~\ref{fig:scan2photRes}(b) shows a comparison of the measured transmission spectrum for $\Omega_p/\Omega_c=0.05$ with the three different models. For all of them the independently measured Rabi frequencies are used as an input for the models and the peak atomic density is estimated similarly to the Autler-Townes measurement and possesses the same uncertainty. Moreover, the dephasings $\gamma_{ge}$ and $\gamma_{gr}$ are the same as for the Autler-Townes measurements. Apparently, all three models fail to describe our measurement.

As shown in Fig.~\ref{fig:scan2photRes}(b), only qualitatively one absorption dip on the negative detuning side is found with the mean-field and MCRE models. However, its position deviates from and cannot be superimposed with the measured one by changing parameters, such as the atomic density, within an acceptable range with respect to the experimental parameters. The stronger absorption predicted by the mean-field model compared to the MCRE simulation stems from the inclusion of the variance $\theta_R$ of the interaction induced shift in the mean-field model, which becomes more important as the fraction $\alpha$ of all blockaded atoms excluding Rydberg
atoms is smaller than $0.12$ for all detunings. This variance is not explicitly included in the MCRE simulation (see App. \ref{app-sec:mean-field_MCRE} for details). 

The low-intensity theory predicts two transmission minima as a consequence of a two-body two-photon resonance. However, even though the assumptions for this model are met, it can not explain the absorption feature on the negative detuning side.

\subsection{Discussion}
Similar to previous experiments \cite{Pritchard:firstEITmot:PRL10,Sevincli:Adams:CPTEIT:JoPB11,Han:ShiftAndDephasing:PRA2016,DeSalvo:MeanField:PRA2016}, the discussed theoretical calculations capture the nonlinear behavior of the measured Autler-Townes spectra for resonant control-laser fields besides the absence of a shift that has already been debated \cite{Sevincli:Adams:CPTEIT:JoPB11,Han:ShiftAndDephasing:PRA2016}. Strikingly, however, all three approaches fail to explain the observed nonlinear absorption spectrum under conditions of EIT. The outlined mean-field description \cite{Han:ShiftAndDephasing:PRA2016} and the MCRE simulations \cite{Gaerttner:REwithPropagation:PRA13} include interaction effects in an approximate way that augments the single-atom equation of motion by a collective level shift produced by surrounding atoms. Consequently, these approaches do not fully account for correlations and entanglement between atoms that arise from strong pairwise interactions in the presence of laser driving. In particular, they neglect pair-state resonances that emerge from strong control-field coupling of blockaded atom pairs \cite{Gaerttner:Resonance:PRL14,Gaul:ResDressing:PRL16,Helmrich:ResDressing:PRL16,Tebben:ResonantEnhancement:PRA19} and lead to enhanced nonlinear absorption around $\Delta\sim\pm\Omega_c/2$. While the presented low-intensity theory \cite{Tebben:ResonantEnhancement:PRA19} exactly accounts for this effect on a two-body level, the comparison to our experiments suggests that the collective influence of multiple interacting atoms plays a significant role for the observed nonlinear absorption spectrum.  

While an exact description of the laser-driven interacting Rydberg atom ensemble is numerically intractable \cite{Gaerttner:REwithPropagation:PRA13} for the large atom number used in our experiment, future improvements of current theoretical approaches may shed light on the discrepancies revealed in this work. For example, a hybrid MCRE scheme, proposed in \cite{Heeg:Evers:HybridModel:PRA12}, combines the described rate-equation description of many interacting atoms with an exact treatment of two-body quantum dynamics. Hereby one identifies close-lying atoms that form otherwise isolated pairs for which the corresponding two-body master equation is solved exactly to obtain a corresponding system of two-body rate equations that yields the exact two-body steady state. This approach is expected to provide an improved description of the nonlinear absorption at low atomic densities \cite{Heeg:Evers:HybridModel:PRA12}. However, the identification of atomic pairs becomes ambiguous at our densities and would hence require a quantum master-equation description of extended atomic clusters in future theoretical work. Moreover, the outlined low-intensity theory for the third-order optical susceptibility of the Rydberg gas \cite{Tebben:ResonantEnhancement:PRA19} may be expanded by truncating the underlying hierarchy of operator equations, discussed in Sec. \ref{sec:chi3}, via a closure relation that takes into account the effect of multiple surrounding Rydberg atoms beyond direct two-body terms. For example, this could be achieved within a systematic cluster expansion and ladder approximation of three-body terms to include the mean-field level-shift generated by Rydberg atoms surrounding a given atomic pair.

\section{Conclusion and Outlook}
\label{sec_concl}
We have experimentally investigated the nonlinear absorption spectrum of a Rydberg EIT medium with strong atomic interactions. Our measurements of the nonlinear behavior of the Autler-Townes absorption peaks for resonant control fields connects to previously observed spectral features, some of which are explainable by existing theories while others are under debate. Investigating the probe beam spectrum on two-photon resonance, i.e., when simultaneously scanning both laser frequencies to maintain EIT conditions, we found significant deviations between an observed broad absorption feature and existing theories. Staying on two-photon resonance with the applied fields implies that any absorption predominantly arises from nonlinear effects such that the presented measurements provide a more stringent test of the theoretical understanding of the underlying optical nonlinearities. The appearance of the observed discrepancies comes as a surprise, in light of the substantial previous investigations of Rydberg-EIT in the semi-classical \cite{Pritchard:firstEITmot:PRL10,Schempp:CPT:PRL10,Sevincli:Adams:CPTEIT:JoPB11,Han:ShiftAndDephasing:PRA2016,DeSalvo:MeanField:PRA2016,Weatherill_2008,Tebben:ResonantEnhancement:PRA19,Sevincli:adibatic:PRL11,Gaul:ResDressing:PRL16,Helmrich:ResDressing:PRL16,Gaerttner:REwithPropagation:PRA13,Gaerttner:Resonance:PRL14,Gaerttner:SemianalyticalModel:PRA14,Li:EITlensing:PRA15} as well as the quantum regime \cite{Pohl:review:16,Dudin:firstQuantumRydEIT:Science12,Peyronel:DissipativeQuantum:Nature2012,Firstenberg:AttractivePhotons:Nature2013,Liang:ThreePhoton:Science2018,Stiesdal:ThreePhotonCorr:PRL18,Cantu:RepulsivePhotons:NatPhys20,Thompson:SymmetryProtectedCollisions:Nature2017,Tiarks:PhotonTransistor:PRL2014,Gorniaczyk:PhotonTransistor:PRL2014,Tiarks:QuantumGate:NatPhys19,Bienias:HighInput:PRR20}. 
The detailed comparison to different and complementary theoretical approaches, presented in this work indeed suggests that an improved treatment of the driven many-body dynamics is necessary to describe EIT in interacting atomic gases. 

\subsection*{ACKNOWLEDGMENTS}
The authors gratefully acknowledge insightful discussions with Valentin Walther and Yong-Chang Zhang. This work is part of and was supported by the DFG Priority Program "GiRyd 1929" (Grant No. DFG WE2661/12-1), and also received support from the DNRF through the Center for Complex Quantum Systems (Grant Agreement No.: DNRF156), the Carlsberg Foundation through the Semper Ardens Research Project QCooL, the EU through the H2020-FETOPEN Grant No. 800942640378 (ErBeStA), the Heidelberg Center for Quantum Dynamics, and Deutsche Forschungsgemeinschaft (DFG, German Research Foundation, Project-ID 273811115, SFB 1225 ISOQUANT). A.T. acknowledges support from the Heidelberg Graduate School for Fundamental Physics. C.H. acknowledges support from the Alexander von Humboldt Foundation.

\appendix

\section{Preparation of the three-level system}
\label{app-sec:exp_details}
$^{87}$Rb atoms are loaded into a magneto-optical trap (MOT) from a high flux cold atom source \cite{Hofmann:setup:Frontiers14}. After a compressed and a dark-MOT phase \cite{Petrich:compressedMOT:JosaB94,Townsend:darkMOT:PRA96} the atoms are transferred into a far detuned, crossed optical dipole trap. Using a combination of optical pumping and Landau Zener transfers between hyperfine sublevels the atoms are prepared in the hyperfine ground state $\ket{g}= \ket{5\text{S}_{1/2},F=2,m_F=2}$ in the presence of a $\SI{30}{G}$ magnetic field. With this procedure we achieve a cigar shaped atom cloud with $1/e^2$ waists of about $40\times40\times\SI{300}{\mu m^3}$, a maximal peak density of $\SI{2e11}{cm^{-3}}$ and a temperature of approximately $\SI{100}{\mu K}$ in a well-defined initial state $\ket{g}$.

The probe and control beams that couple the ground to the Rydberg state in a two-photon process via the intermediate state are right- and left-circularly polarized, respectively, with respect to the applied magnetic field. In combination with the careful preparation of the ground state, this ensures that our EIT setup is realized as a well-defined three-level-ladder system. Moreover, both lasers are locked to a stable, high-finesse Fabry-P\'{e}rot cavity with a free spectral range of $\SI{1.5}{GHz}$,  allowing for detunings $\Delta_{p,c}$ of the two beams of up to $\SI{750}{MHz}$ and small laser linewidths below $\SI{10}{kHz}$.


\section{Characterization of dephasing in the two-level system}
\label{app-sec:2lvl_dephasing}
\begin{figure}[t!]
	\includegraphics[width=0.8\linewidth]{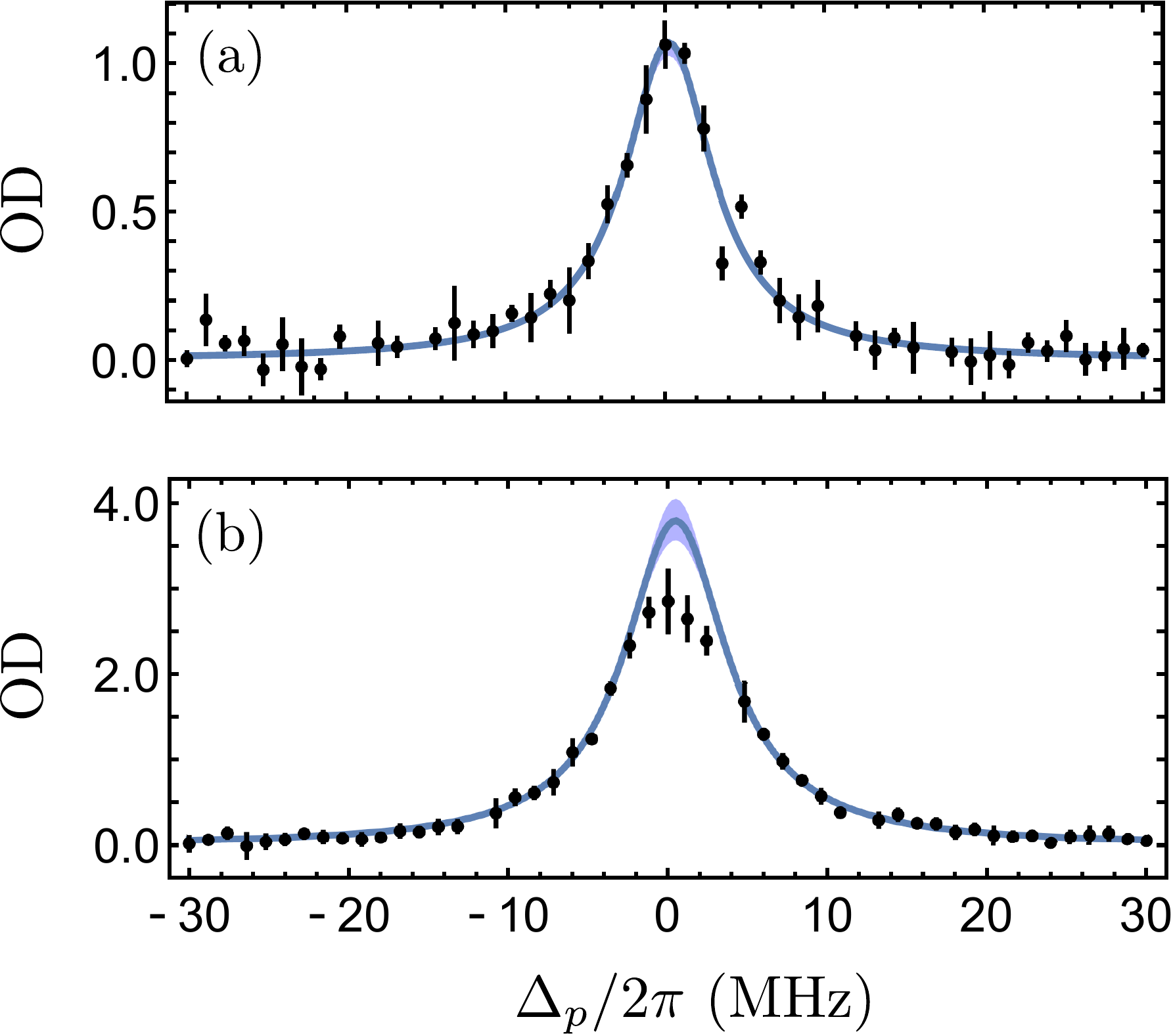}
	\caption{Optical depth equal to $-\ln{(T)}$ as a function of the probe beam detuning $\Delta_p$ in the absence of the control beam ($\Omega_c=0$) and for peak atomic densities of (a) $\rho_0\approx\SI{0.02}{\mu m^{-3}}$ and (b) $\rho_0\approx\SI{0.2}{\mu m^{-3}}$. The result of the mean-field model is shown as a solid line, with the shaded area indicating the uncertainty in the determined dephasing rate $\gamma_{ge}$.}
	\label{fig:2lvl_scan} 
\end{figure}
In order to determine the dephasing $\gamma_{ge}$ of the excited state coherence, we measure the optical depth ($\text{OD}$) of the atomic cloud as a function of the probe beam detuning $\Delta_p$ in the absence of the control beam ($\Omega_c=0$). For comparability with our measurements in the main text, we choose a rather high probe Rabi frequency $\Omega_p/2\pi=\SI{1.7}{MHz}$.

In the low-density regime, shown in Fig.~\ref{fig:2lvl_scan}(a), we extract a linewidth of $2\pi\times\SI[separate-uncertainty]{6.8(3)}{MHz}$ and a peak OD of $\SI[separate-uncertainty]{1.06(6)}{}$ using a Lorentzian fit to the data, where the error is the weighted error from the fit. In the high density regime, as depicted in Fig.~\ref{fig:2lvl_scan}(b), we exclude data points with an $\text{OD}$ grater than $1.8$ from the fitting routine and obtain a rather large linewidth of $2\pi\times\SI[separate-uncertainty]{7.9(5)}{MHz}$ with a peak OD of $\SI[separate-uncertainty]{3.8(2)}{}$. From the fit results and the independently measured geometry of the atomic cloud we extract numerically using the mean-field model the corresponding dephasing rates $\gamma_{ge}/2\pi=\SI[separate-uncertainty]{0.5(3)}{MHz}$ and $\gamma_{ge}/2\pi=\SI[separate-uncertainty]{1.7(5)}{MHz}$ for the low- and high-density measurements, respectively. The result is shown as blue solid lines in Fig.~\ref{fig:2lvl_scan}.

The natural linewidth due to population decay, power broadening, and a reduction of the linewidth due to propagation effects is intrinsically included in the mean-field model such that these effects cannot be the source of the observed dephasing. The Doppler shift for rubidium atoms at a temperature of $\SI{100}{\mu K}$ is approximately $\SI{200}{kHz}$ such that Doppler broadening is negligible compared to the strong dephasing observed. Density-dependent dephasing mechanisms that could cause such a broadening are atomic collisions and rescattering of photons. Estimating the broadening due to collisions \cite{Hertel:AMOPhysics:Springer15} by calculating the collision rate from the atomic velocity and the mean-free-path shows that the temperature or the density of the atomic gas is too low to explain this large amount of dephasing. However, due to the large extent of the atomic cloud transversal to the propagation direction, the transverse optical depth is large and allows for multiple rescattering of the photons \cite{Labeyrie:SlowDiffusionLight:PRL03,Labeyrie:RadiationTrapping:APB05}, which can broaden the line at high densities.

For the theoretical curves presented in Figs.~\ref{fig:AutlerTownes} and~\ref{fig:scan2photRes}(b) we extrapolate the dephasings determined here based on the independently measured optical depth for the EIT spectra. For an estimation of the atomic density in these measurements we fit the Autler-Townes measurement in the noninteracting regime with the peak atomic density as the only free fitting parameter. Its uncertainty results from shot-to-shot fluctuations of $\pm8\%$, a statistical error of $\pm2\%$ given by the uncertainty of the deduced dephasing rate $\gamma_{ge}$ and a systematic overestimation of the propagation length resulting in an error of $+4\%$ for the atomic density. The uncertainty in the density is included in Figs.~\ref{fig:AutlerTownes} and~\ref{fig:scan2photRes}(b) as shaded areas.

\section{Rydberg excitation measurement on two-photon resonance}
\label{app-sec:ion_measurement}
Besides a transmission measurement, our setup also allows us to detect Rydberg excitations, that remain in the atomic cloud after turning off the EIT lasers, by field ionization of the atom cloud and subsequent detection of the resulting ions on a micro-channel plate (MCP). Here, the detection efficiency, measured by depletion imaging \cite{Ferreira:DepletionImaging:JPhysB20}, is about $0.10$ ions per Rydberg excitation and in the absence of the control beam, where no ions can be created, we measure $\SI[separate-uncertainty=true]{0.743(96)}{}$ counts, where the error is the standard error of the mean, setting a threshold for the detection of Rydberg excitations.

\begin{figure}[t!]
	\includegraphics[width=0.95\linewidth]{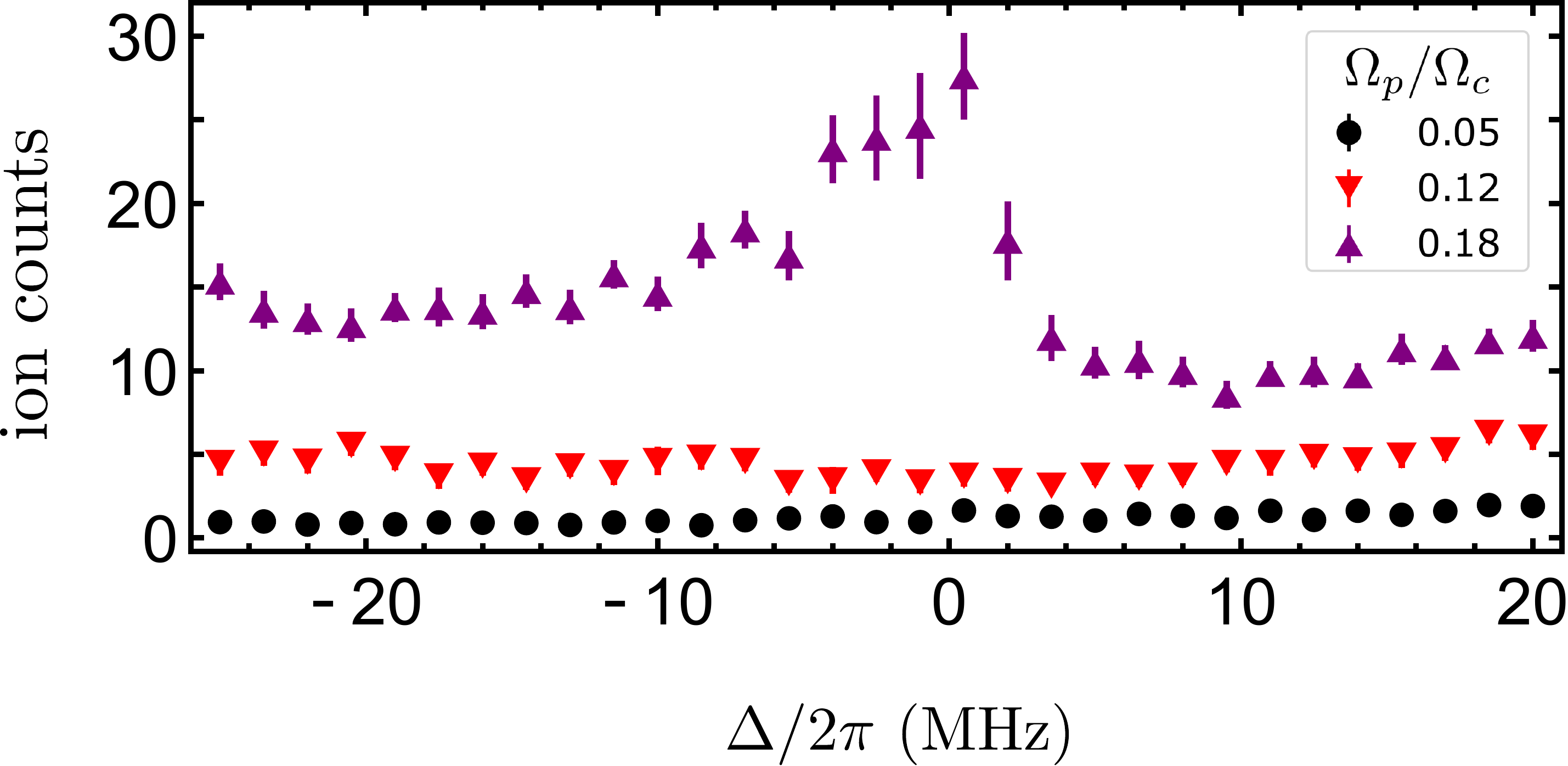}
	\caption{Measurements of ion counts on two-photon resonance ($\delta=0$) for different ratios $\Omega_p/\Omega_c$. The ions were detected simultaneously with the measurement of Fig.~\ref{fig:scan2photRes}. }
	\label{fig:scan2photRes_ions} 
\end{figure}

For the measurement on two-photon resonance presented in Fig.~\ref{fig:scan2photRes}(a) we simultaneously recorded the ion counts on the MCP, as shown in Fig.~\ref{fig:scan2photRes_ions}. For the ratio of $\Omega_p/\Omega_c=0.05$, where the nonlinear absorption in the transmission measurements appears, the number of detected ions is about $1$ for all detunings, as shown by the black circles in Fig.~\ref{fig:scan2photRes_ions}. When increasing the ratio of the two Rabi frequencies further, the ion count increases to about $4$ counts, but stays approximately constant over the whole range of single-photon detunings. For the highest measured ratio the number of detected ions increases significantly with a maximum around zero single-photon detuning.

In the case of a coherent evolution in the EIT system we would expect to detect no ions after the EIT sequence. Therefore, we attribute the observed ions at high ratios of the two Rabi frequencies to stationary Rydberg excitations in the medium. As these excitations do not get depumped by the control beam into the decaying intermediate state it is to be presumed that these are excitations in other than the $\ket{48\text{S}_{1/2}, m_j=1/2}$ Rydberg state. An explanation for the creation of theses excitations might be radiation trapping \cite{Labeyrie:SlowDiffusionLight:PRL03,Sadler:RadiationTrappingPollutans:PRA17} and subsequent state-changing collisions or antiblockade excitation of Rydberg states that are not coupled by the control laser. These unwanted Rydberg excitations have already been observed and termed Rydberg pollutants in ref. \cite{Bienias:HighInput:PRR20}.

For the atomic density and the ratio of the two Rabi frequencies used for the interacting Autler-Townes measurement presented in Fig.\ref{fig:AutlerTownes}(b), which connect to the red curve in Fig.~\ref{fig:scan2photRes_ions}, the ion measurement suggests the presence of a large number of Rydberg excitations that could lead to the observed slightly lower transmission around resonance. 

However, for the measurement at $\Omega_p/\Omega_c\approx 0.05$ we measure approximately one ion count at the position of the transmission dip at $\Delta_\text{min}=\SI{-8}{MHz}$. According to the detection efficiency, this sets an upper bound of $10$ on the number $N_\text{Ryd}$ of Rydberg excitations in the medium. We now consider a worst case scenario in order to estimate the maximal influence of these excitations on the probe beam propagation. For this purpose we assume, that all these excitations are located in the integrated region of $4\times4$ pixels, where the probe beam transmission is evaluated. Furthermore, we assume that these excitations are atoms in the $48$P Rydberg state that process strong dipolar interactions with the $48$S Rydberg state with a coefficient $c_3$ of about $\SI{1.7}{GHz\mu m^3}$. 

We estimate the resulting absorption from theses excitations as follows. 
Each Rydberg excitation renders the medium absorptive in a spherical volume given by the blockade radius $R_b$, which is approximately $\SI{3.1}{\mu m}$ at $\Delta_\text{min}$.
Assuming that all excitations are placed in a chain behind each other the resulting optical depth $\text{OD}_\text{Ryd}=\text{OD}_\text{off}2R_bN_\text{Ryd} /L$ can be calculated from the propagation distance $L$ through the whole atomic cloud, the peak atomic density $\rho_0$, which is given in the caption of Fig.~\ref{fig:scan2photRes}, and the off-resonant optical depth $\text{OD}_\text{off}=\sigma_\text{off}\rho_0L$ of two-level atoms. Here the off-resonant cross section $\sigma_\text{off}=a\sigma_0$ is the resonant cross section $\sigma_0$ multiplied by a factor $a=0.126$ that takes into account the Lorentzian lineshape of the two-level absorption with decay rate $\gamma_e$. In the last step we have to account for the fact, that the transversal size $A_\text{Ryd}=\pi R_b^2$ of one blockaded volume is smaller than the evaluated pixel area $A=(4\times\SI{2.2}{\mu m})^2$ on the CCD camera and use the scale $s=A_\text{Ryd}/A$ to finally obtain the transmission $T_\text{Ryd}=(1-s)+s\exp(-\text{OD}_\text{Ryd})\approx 0.87$ in the presence of ten Rydberg excitations.

Overall, this estimation in a worst case scenario results in an upper bound of $13\%$ for the probe beam absorption solely due to these Rydberg excitations. Therefore, for the ratio of $\Omega_p/\Omega_c=0.05$, unwanted Rydberg excitations cannot explain the observed strong absorption feature.

\section{Comparison of mean-field and MCRE model}
\label{app-sec:mean-field_MCRE}

In the spectra on two-photon resonance a stronger absorption is predicted by the mean-field model than by the result of the Monte Carlo rate-equation model, shown in Fig.~\ref{fig:scan2photRes}(b). We explain in the following that this results from the assumption of how the interaction-induced level shift is included in the two models.

On one hand, in the MCRE model, the total interaction-induced level shift $\Delta_\text{int}^{(i)}$ experienced by an atom $i$ is determined by the sum $\sum_{j\neq i}\Delta_{ij}=\sum_{j\neq i}C_6/r_{ij}^6$ over all shifts induced by the surrounding Rydberg atoms \cite{Gaerttner:REwithPropagation:PRA13}. As the MCRE simulation is seeded with a distribution of atoms according to the geometry of the experiment, the inter-atomic distances $r_{ij}$ vary, which immediately leads to a certain variation of the level shifts $\Delta_{ij}$.

\begin{figure*}[t!]
	\includegraphics[width=0.85\linewidth]{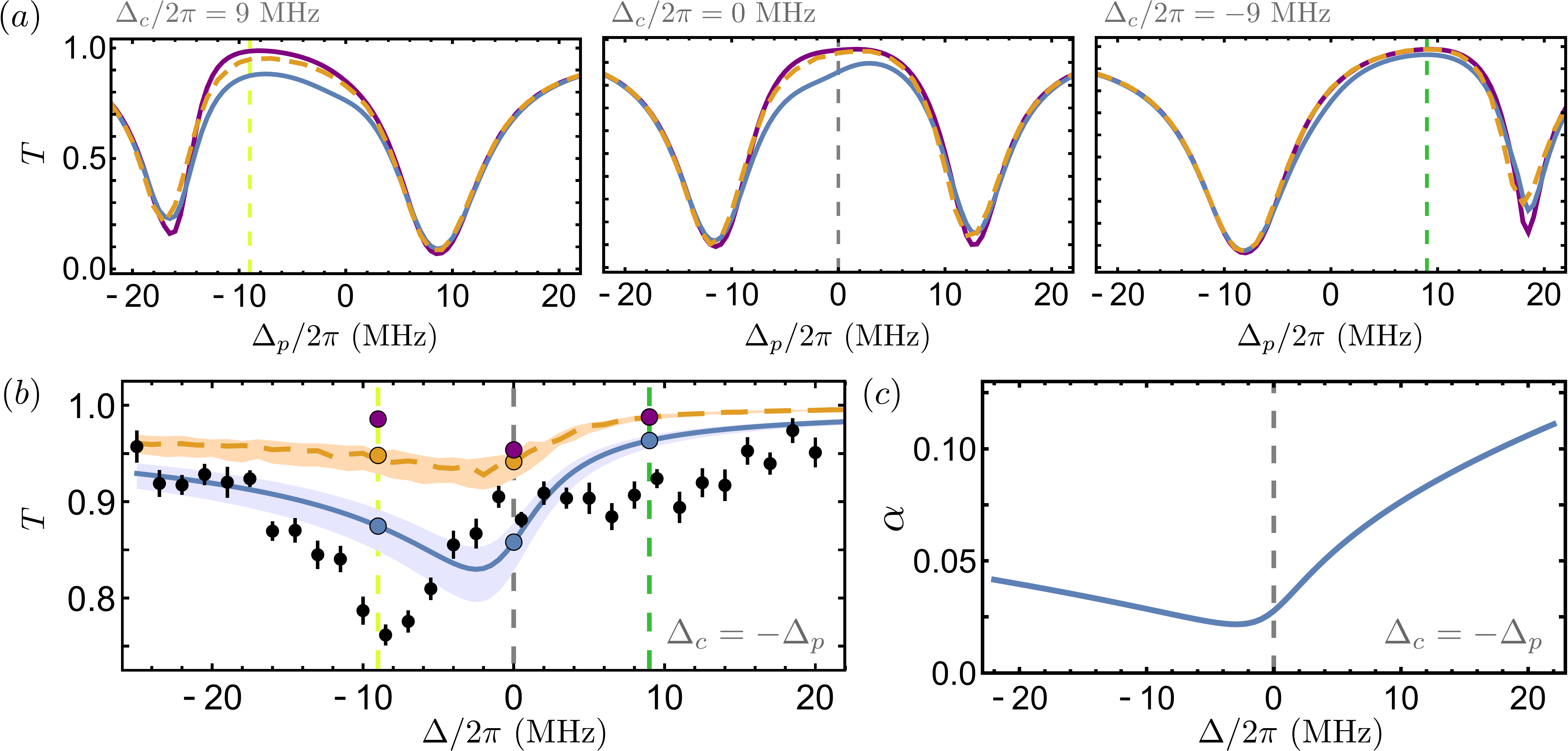}
	\caption{Comparison of the mean-field model [for $\theta_R\neq 0$ (blue solid lines) and $\theta_R=0$ (purple solid lines)] and MCRE simulation (orange dashed lines) for the parameters of Fig.~\ref{fig:scan2photRes}(b). (a) Autler-Townes transmission spectra as a function of the probe beam detuning $\Delta_p$ for different control beam detunings $\Delta_c/2\pi=\left\{9,0,-9\right\}$ MHz. (b) Transmission $T$ against the single-photon detuning $\Delta$ on two-photon resonance ($\Delta_c=-\Delta_p$), as in Fig.~\ref{fig:scan2photRes}(b). Black points depict the measured spectrum and the results of the two theoretical models are shown. Colored circles indicate the transmission values of the corresponding Autler-Townes spectrum in (a). (c) Plot of $\alpha$ of the mean-field model against the single-photon detuning $\Delta$. For all theoretical curves the parameters are the same as in Fig.~\ref{fig:scan2photRes}(b). For a discussion of the curves see the main text.}
	\label{fig:comp_mean-field_MCRE} 
\end{figure*}
On the other hand, the considered mean-field model is based on the noninteracting single-body susceptibility and includes an interaction-induced level shift therein \cite{Han:ShiftAndDephasing:PRA2016}. This means that nothing like an atomic distribution, and therefore no variance of the level shift is considered \textit{a priori}. For distances smaller than the blockade radius, the level shift is completely determined by an integration over the radius $r$ and the resulting susceptibility $\chi_B$ inside the blockade radius is therefore unambiguously defined. However, for the susceptibility $\chi_E$ outside the blockaded sphere, assumptions about the average level shift $\Delta_R$ and its variance $\theta_R$ have to be made. \citet{Han:ShiftAndDephasing:PRA2016} calculated both based on a mean-field assumption as well as on an assumption for the Rydberg excitation fraction. Explicitly, the formula for the variance reads \cite{Han:ShiftAndDephasing:PRA2016} 
\begin{equation}
\theta_R\approx\int_{R_b}^\infty f_R\rho_0\frac{C_6^2}{r^{12}}4\pi r^2dr=\frac{4\pi C_6^2f_R\rho_0}{9R_b^9}\,, 
\end{equation}
where $f_R=\frac{f_0}{1-f_0+f_0\rho_0V_B}$ \cite{Han:ShiftAndDephasing:PRA2016} is the Rydberg excitation fraction in the interacting regime and $f_0$ the one in the non-interacting regime. Here $V_B$ is the spherical volume spanned by one blockade radius $R_B$.
The variance is then included as an effective dephasing $\gamma_{gr}=\sqrt{\theta_R}$ of the Rydberg coherence in the single-body susceptibility \cite{Han:ShiftAndDephasing:PRA2016}.
Finally, the Rydberg excitation fraction determines the weight between the two parts of the overall model susceptibility $\bar\chi=\alpha\chi_B+(1-\alpha)\chi_E$, as $\alpha\propto f_R$ \cite{Han:ShiftAndDephasing:PRA2016}. This implies, that the more one enters the blockaded regime (large $\alpha$), the less weight is put on the assumption made for the variance of the level-shift entering $\chi_E$.

For the Autler-Townes measurement in the interacting regime, $\alpha$ is larger than $0.6$ around the Autler-Townes transmission minima. Hence, the contributions of $\chi_B$ and $\chi_E$ are quite similar such that the relative importance of including a variance of the level shift is small. As a result, the mean-field and MCRE models give similar predictions for the transmission spectrum. Only around resonance, where $\alpha$ is about $0.2$, deviations between the two models start to appear.

For the measurement on two-photon resonance, $\alpha<0.12$ for all single-photon detunings $\Delta$ and is especially only about $0.025$ at $\Delta/2\pi=\SI{-2}{MHz}$, as shown in Fig.~\ref{fig:comp_mean-field_MCRE}(c). At the same detuning the effective dephasing on the Rydberg coherence $\sqrt{\theta_R}/2\pi$ is as large as $\SI{5}{MHz}$, putting a great deal of weight on the assumptions made in the mean-field model.

In order to show that the transmission curve on two-photon resonance predicted by the mean-field model is dominated by the variance $\theta_R$, we show in Fig.~\ref{fig:comp_mean-field_MCRE}(a) the Autler-Townes transmission spectra for three different detunings $\Delta_c/2\pi=\left\{9,0,-9\right\}$ MHz corresponding to the points marked on the curves on two-photon resonance depicted in Fig.~\ref{fig:comp_mean-field_MCRE}(b). The mean-field model with (without, $\theta_R=0$) the variance $\theta_R$ is shown in blue (purple) and the result of the MCRE simulation is shown in orange for comparison. 

For positive single-photon detunings $\Delta>0$, where $\alpha$ is larger, the mean-field model and the MCRE model almost agree for the Autler-Townes spectrum with $\Delta_c/2\pi=\SI{-9}{MHz}$ and setting $\theta_R=0$ makes them almost identical. For zero and negative single-photon detunings, $\alpha$ is smaller and a deviation between the mean-field model and the MCRE model is apparent. Completely excluding the variance of the interaction-induced level shift ($\theta_R=0$) lets the result of the two models become similar, but seems to underestimate the absorption compared to the MCRE model. This highlights the importance of including a spatial variation of the level shift in an appropriate manner.

Overall, the considered mean-field model is dominated by the variance $\theta_R$ of the average interaction-induced level shift, whenever $\alpha$ is small, which is the case when the fraction of blockaded atoms excluding Rydberg excitations is small. In this regime, in which our two-photon measurement mainly belongs, deviations between the mean-field model and the MCRE simulation become apparent.


\begin{thebibliography}{46}%
	\makeatletter
	\providecommand \@ifxundefined [1]{%
		\@ifx{#1\undefined}
	}%
	\providecommand \@ifnum [1]{%
		\ifnum #1\expandafter \@firstoftwo
		\else \expandafter \@secondoftwo
		\fi
	}%
	\providecommand \@ifx [1]{%
		\ifx #1\expandafter \@firstoftwo
		\else \expandafter \@secondoftwo
		\fi
	}%
	\providecommand \natexlab [1]{#1}%
	\providecommand \enquote  [1]{``#1''}%
	\providecommand \bibnamefont  [1]{#1}%
	\providecommand \bibfnamefont [1]{#1}%
	\providecommand \citenamefont [1]{#1}%
	\providecommand \href@noop [0]{\@secondoftwo}%
	\providecommand \href [0]{\begingroup \@sanitize@url \@href}%
	\providecommand \@href[1]{\@@startlink{#1}\@@href}%
	\providecommand \@@href[1]{\endgroup#1\@@endlink}%
	\providecommand \@sanitize@url [0]{\catcode `\\12\catcode `\$12\catcode
		`\&12\catcode `\#12\catcode `\^12\catcode `\_12\catcode `\%12\relax}%
	\providecommand \@@startlink[1]{}%
	\providecommand \@@endlink[0]{}%
	\providecommand \url  [0]{\begingroup\@sanitize@url \@url }%
	\providecommand \@url [1]{\endgroup\@href {#1}{\urlprefix }}%
	\providecommand \urlprefix  [0]{URL }%
	\providecommand \Eprint [0]{\href }%
	\providecommand \doibase [0]{http://dx.doi.org/}%
	\providecommand \selectlanguage [0]{\@gobble}%
	\providecommand \bibinfo  [0]{\@secondoftwo}%
	\providecommand \bibfield  [0]{\@secondoftwo}%
	\providecommand \translation [1]{[#1]}%
	\providecommand \BibitemOpen [0]{}%
	\providecommand \bibitemStop [0]{}%
	\providecommand \bibitemNoStop [0]{.\EOS\space}%
	\providecommand \EOS [0]{\spacefactor3000\relax}%
	\providecommand \BibitemShut  [1]{\csname bibitem#1\endcsname}%
	\let\auto@bib@innerbib\@empty
	\bibitem [{\citenamefont {Firstenberg}\ \emph {et~al.}(2016)\citenamefont
		{Firstenberg}, \citenamefont {Adams},\ and\ \citenamefont
		{Hofferberth}}]{FirstenbergAdamsHofferberth:review:JPhysB2016}%
	\BibitemOpen
	\bibfield  {author} {\bibinfo {author} {\bibfnamefont {O.}~\bibnamefont
			{Firstenberg}}, \bibinfo {author} {\bibfnamefont {C.~S.}\ \bibnamefont
			{Adams}}, \ and\ \bibinfo {author} {\bibfnamefont {S.}~\bibnamefont
			{Hofferberth}},\ }\href {\doibase 10.1088/0953-4075/49/15/152003} {\bibfield
		{journal} {\bibinfo  {journal} {J. Phys. B}\ }\textbf {\bibinfo {volume}
			{49}},\ \bibinfo {pages} {152003} (\bibinfo {year} {2016})}\BibitemShut
	{NoStop}%
%
	\bibitem [{\citenamefont {Murray}\ and\ \citenamefont
		{Pohl}(2016)}]{Pohl:review:16}%
	\BibitemOpen
	\bibfield  {author} {\bibinfo {author} {\bibfnamefont {C.}~\bibnamefont
			{Murray}}\ and\ \bibinfo {author} {\bibfnamefont {T.}~\bibnamefont {Pohl}}\
	}\href {\doibase 10.1016/bs.aamop.2016.04.005} {\bibfield
	{journal} {\bibinfo  {journal} {Adv. At. Mol. Opt. Phys.}\ }\textbf {\bibinfo {volume}
		{65}},\ \bibinfo {pages} {321} (\bibinfo {year} {2016})}
	\BibitemShut {NoStop}%
%
	\bibitem [{\citenamefont {Firstenberg}\ \emph {et~al.}(2013)\citenamefont
		{Firstenberg}, \citenamefont {Peyronel}, \citenamefont {Liang}, \citenamefont
		{Gorshkov}, \citenamefont {Lukin},\ and\ \citenamefont
		{Vuleti{\'c}}}]{Firstenberg:AttractivePhotons:Nature2013}%
	\BibitemOpen
	\bibfield  {author} {\bibinfo {author} {\bibfnamefont {O.}~\bibnamefont
			{Firstenberg}}, \bibinfo {author} {\bibfnamefont {T.}~\bibnamefont
			{Peyronel}}, \bibinfo {author} {\bibfnamefont {Q.-Y.}\ \bibnamefont {Liang}},
		\bibinfo {author} {\bibfnamefont {A.~V.}\ \bibnamefont {Gorshkov}}, \bibinfo
		{author} {\bibfnamefont {M.~D.}\ \bibnamefont {Lukin}}, \ and\ \bibinfo
		{author} {\bibfnamefont {V.}~\bibnamefont {Vuleti{\'c}}},\ }\href {\doibase
		10.1038/nature12512} {\bibfield  {journal} {\bibinfo  {journal} {Nature (London)}\
		}\textbf {\bibinfo {volume} {502}},\ \bibinfo {pages} {71} (\bibinfo {year}
		{2013})}\BibitemShut {NoStop}%
	\bibitem [{\citenamefont {Otterbach}\ \emph {et~al.}(2013)\citenamefont
		{Otterbach}, \citenamefont {Moos}, \citenamefont {Muth},\ and\ \citenamefont
		{Fleischhauer}}]{Otterbach:WignerCrystal:PRL13}%
	\BibitemOpen
	\bibfield  {author} {\bibinfo {author} {\bibfnamefont {J.}~\bibnamefont
			{Otterbach}}, \bibinfo {author} {\bibfnamefont {M.}~\bibnamefont {Moos}},
		\bibinfo {author} {\bibfnamefont {D.}~\bibnamefont {Muth}}, \ and\ \bibinfo
		{author} {\bibfnamefont {M.}~\bibnamefont {Fleischhauer}},\ }\href {\doibase
		10.1103/PhysRevLett.111.113001} {\bibfield  {journal} {\bibinfo  {journal}
			{Phys. Rev. Lett.}\ }\textbf {\bibinfo {volume} {111}},\ \bibinfo {pages}
		{113001} (\bibinfo {year} {2013})}\BibitemShut {NoStop}%
	\bibitem [{\citenamefont {Bienias}\ \emph {et~al.}(2014)\citenamefont
		{Bienias}, \citenamefont {Choi}, \citenamefont {Firstenberg}, \citenamefont
		{Maghrebi}, \citenamefont {Gullans}, \citenamefont {Lukin}, \citenamefont
		{Gorshkov},\ and\ \citenamefont
		{B\"uchler}}]{Bienias:ScatteringResonances:PRA2014}%
	\BibitemOpen
	\bibfield  {author} {\bibinfo {author} {\bibfnamefont {P.}~\bibnamefont
			{Bienias}}, \bibinfo {author} {\bibfnamefont {S.}~\bibnamefont {Choi}},
		\bibinfo {author} {\bibfnamefont {O.}~\bibnamefont {Firstenberg}}, \bibinfo
		{author} {\bibfnamefont {M.~F.}\ \bibnamefont {Maghrebi}}, \bibinfo {author}
		{\bibfnamefont {M.}~\bibnamefont {Gullans}}, \bibinfo {author} {\bibfnamefont
			{M.~D.}\ \bibnamefont {Lukin}}, \bibinfo {author} {\bibfnamefont {A.~V.}\
			\bibnamefont {Gorshkov}}, \ and\ \bibinfo {author} {\bibfnamefont {H.~P.}\
			\bibnamefont {B\"uchler}},\ }\href {\doibase 10.1103/PhysRevA.90.053804}
	{\bibfield  {journal} {\bibinfo  {journal} {Phys. Rev. A}\ }\textbf {\bibinfo
			{volume} {90}},\ \bibinfo {pages} {053804} (\bibinfo {year}
		{2014})}\BibitemShut {NoStop}%
	\bibitem [{\citenamefont {Moos}\ \emph {et~al.}(2015)\citenamefont {Moos},
		\citenamefont {H\"oning}, \citenamefont {Unanyan},\ and\ \citenamefont
		{Fleischhauer}}]{Moos:ManyBodyPolaritons:PRA15}%
	\BibitemOpen
	\bibfield  {author} {\bibinfo {author} {\bibfnamefont {M.}~\bibnamefont
			{Moos}}, \bibinfo {author} {\bibfnamefont {M.}~\bibnamefont {H\"oning}},
		\bibinfo {author} {\bibfnamefont {R.}~\bibnamefont {Unanyan}}, \ and\
		\bibinfo {author} {\bibfnamefont {M.}~\bibnamefont {Fleischhauer}},\ }\href
	{\doibase 10.1103/PhysRevA.92.053846} {\bibfield  {journal} {\bibinfo
			{journal} {Phys. Rev. A}\ }\textbf {\bibinfo {volume} {92}},\ \bibinfo
		{pages} {053846} (\bibinfo {year} {2015})}\BibitemShut {NoStop}%
	\bibitem [{\citenamefont {Carusotto}\ and\ \citenamefont
		{Ciuti}(2013)}]{Carusotto:QuantumFluid:RMP13}%
	\BibitemOpen
	\bibfield  {author} {\bibinfo {author} {\bibfnamefont {I.}~\bibnamefont
			{Carusotto}}\ and\ \bibinfo {author} {\bibfnamefont {C.}~\bibnamefont
			{Ciuti}},\ }\href {\doibase 10.1103/RevModPhys.85.299} {\bibfield  {journal}
		{\bibinfo  {journal} {Rev. Mod. Phys.}\ }\textbf {\bibinfo {volume} {85}},\
		\bibinfo {pages} {299} (\bibinfo {year} {2013})}\BibitemShut {NoStop}%
	\bibitem [{\citenamefont {Chang}\ \emph {et~al.}(2014)\citenamefont {Chang},
		\citenamefont {Vuleti{\'c}},\ and\ \citenamefont
		{Lukin}}]{Chang:QuantumOptics:NatPhot14}%
	\BibitemOpen
	\bibfield  {author} {\bibinfo {author} {\bibfnamefont {D.~E.}\ \bibnamefont
			{Chang}}, \bibinfo {author} {\bibfnamefont {V.}~\bibnamefont {Vuleti{\'c}}},
		\ and\ \bibinfo {author} {\bibfnamefont {M.~D.}\ \bibnamefont {Lukin}},\
	}\href {\doibase 10.1038/nphoton.2014.192} {\bibfield  {journal} {\bibinfo
			{journal} {Nature Photonics}\ }\textbf {\bibinfo {volume} {8}},\ \bibinfo
		{pages} {685} (\bibinfo {year} {2014})}\BibitemShut {NoStop}%
	\bibitem [{\citenamefont {Dudin}\ and\ \citenamefont
		{Kuzmich}(2012)}]{Dudin:firstQuantumRydEIT:Science12}%
	\BibitemOpen
	\bibfield  {author} {\bibinfo {author} {\bibfnamefont {Y.~O.}\ \bibnamefont
			{Dudin}}\ and\ \bibinfo {author} {\bibfnamefont {A.}~\bibnamefont
			{Kuzmich}},\ }\href {\doibase 10.1126/science.1217901} {\bibfield  {journal}
		{\bibinfo  {journal} {Science}\ }\textbf {\bibinfo {volume} {336}},\ \bibinfo
		{pages} {887} (\bibinfo {year} {2012})}\BibitemShut {NoStop}%
	\bibitem [{\citenamefont {Peyronel}\ \emph {et~al.}(2012)\citenamefont
		{Peyronel}, \citenamefont {Firstenberg}, \citenamefont {Liang}, \citenamefont
		{Hofferberth}, \citenamefont {Gorshkov}, \citenamefont {Pohl}, \citenamefont
		{Lukin},\ and\ \citenamefont
		{Vuleti{\'c}}}]{Peyronel:DissipativeQuantum:Nature2012}%
	\BibitemOpen
	\bibfield  {author} {\bibinfo {author} {\bibfnamefont {T.}~\bibnamefont
			{Peyronel}}, \bibinfo {author} {\bibfnamefont {O.}~\bibnamefont
			{Firstenberg}}, \bibinfo {author} {\bibfnamefont {Q.-Y.}\ \bibnamefont
			{Liang}}, \bibinfo {author} {\bibfnamefont {S.}~\bibnamefont {Hofferberth}},
		\bibinfo {author} {\bibfnamefont {A.~V.}\ \bibnamefont {Gorshkov}}, \bibinfo
		{author} {\bibfnamefont {T.}~\bibnamefont {Pohl}}, \bibinfo {author}
		{\bibfnamefont {M.~D.}\ \bibnamefont {Lukin}}, \ and\ \bibinfo {author}
		{\bibfnamefont {V.}~\bibnamefont {Vuleti{\'c}}},\ }\href {\doibase
		10.1038/nature11361} {\bibfield  {journal} {\bibinfo  {journal} {Nature (London)}\
		}\textbf {\bibinfo {volume} {488}},\ \bibinfo {pages} {57} (\bibinfo {year}
		{2012})}\BibitemShut {NoStop}%
	\bibitem [{\citenamefont {Liang}\ \emph {et~al.}(2018)\citenamefont {Liang},
		\citenamefont {Venkatramani}, \citenamefont {Cantu}, \citenamefont
		{Nicholson}, \citenamefont {Gullans}, \citenamefont {Gorshkov}, \citenamefont
		{Thompson}, \citenamefont {Chin}, \citenamefont {Lukin},\ and\ \citenamefont
		{Vuleti{\'c}}}]{Liang:ThreePhoton:Science2018}%
	\BibitemOpen
	\bibfield  {author} {\bibinfo {author} {\bibfnamefont {Q.-Y.}\ \bibnamefont
			{Liang}}, \bibinfo {author} {\bibfnamefont {A.~V.}\ \bibnamefont
			{Venkatramani}}, \bibinfo {author} {\bibfnamefont {S.~H.}\ \bibnamefont
			{Cantu}}, \bibinfo {author} {\bibfnamefont {T.~L.}\ \bibnamefont
			{Nicholson}}, \bibinfo {author} {\bibfnamefont {M.~J.}\ \bibnamefont
			{Gullans}}, \bibinfo {author} {\bibfnamefont {A.~V.}\ \bibnamefont
			{Gorshkov}}, \bibinfo {author} {\bibfnamefont {J.~D.}\ \bibnamefont
			{Thompson}}, \bibinfo {author} {\bibfnamefont {C.}~\bibnamefont {Chin}},
		\bibinfo {author} {\bibfnamefont {M.~D.}\ \bibnamefont {Lukin}}, \ and\
		\bibinfo {author} {\bibfnamefont {V.}~\bibnamefont {Vuleti{\'c}}},\ }\href
	{\doibase 10.1126/science.aao7293} {\bibfield  {journal} {\bibinfo  {journal}
			{Science}\ }\textbf {\bibinfo {volume} {359}},\ \bibinfo {pages} {783}
		(\bibinfo {year} {2018})}\BibitemShut {NoStop}%
	\bibitem [{\citenamefont {Stiesdal}\ \emph {et~al.}(2018)\citenamefont
		{Stiesdal}, \citenamefont {Kumlin}, \citenamefont {Kleinbeck}, \citenamefont
		{Lunt}, \citenamefont {Braun}, \citenamefont {Paris-Mandoki}, \citenamefont
		{Tresp}, \citenamefont {B\"uchler},\ and\ \citenamefont
		{Hofferberth}}]{Stiesdal:ThreePhotonCorr:PRL18}%
	\BibitemOpen
	\bibfield  {author} {\bibinfo {author} {\bibfnamefont {N.}~\bibnamefont
			{Stiesdal}}, \bibinfo {author} {\bibfnamefont {J.}~\bibnamefont {Kumlin}},
		\bibinfo {author} {\bibfnamefont {K.}~\bibnamefont {Kleinbeck}}, \bibinfo
		{author} {\bibfnamefont {P.}~\bibnamefont {Lunt}}, \bibinfo {author}
		{\bibfnamefont {C.}~\bibnamefont {Braun}}, \bibinfo {author} {\bibfnamefont
			{A.}~\bibnamefont {Paris-Mandoki}}, \bibinfo {author} {\bibfnamefont
			{C.}~\bibnamefont {Tresp}}, \bibinfo {author} {\bibfnamefont {H.~P.}\
			\bibnamefont {B\"uchler}}, \ and\ \bibinfo {author} {\bibfnamefont
			{S.}~\bibnamefont {Hofferberth}},\ }\href {\doibase
		10.1103/PhysRevLett.121.103601} {\bibfield  {journal} {\bibinfo  {journal}
			{Phys. Rev. Lett.}\ }\textbf {\bibinfo {volume} {121}},\ \bibinfo {pages}
		{103601} (\bibinfo {year} {2018})}\BibitemShut {NoStop}%
	\bibitem [{\citenamefont {Cantu}\ \emph {et~al.}(2020)\citenamefont {Cantu},
		\citenamefont {Venkatramani}, \citenamefont {Xu}, \citenamefont {Zhou},
		\citenamefont {Jelenkovi{\'c}}, \citenamefont {Lukin},\ and\ \citenamefont
		{Vuleti{\'c}}}]{Cantu:RepulsivePhotons:NatPhys20}%
	\BibitemOpen
	\bibfield  {author} {\bibinfo {author} {\bibfnamefont {S.~H.}\ \bibnamefont
			{Cantu}}, \bibinfo {author} {\bibfnamefont {A.~V.}\ \bibnamefont
			{Venkatramani}}, \bibinfo {author} {\bibfnamefont {W.}~\bibnamefont {Xu}},
		\bibinfo {author} {\bibfnamefont {L.}~\bibnamefont {Zhou}}, \bibinfo {author}
		{\bibfnamefont {B.}~\bibnamefont {Jelenkovi{\'c}}}, \bibinfo {author}
		{\bibfnamefont {M.~D.}\ \bibnamefont {Lukin}}, \ and\ \bibinfo {author}
		{\bibfnamefont {V.}~\bibnamefont {Vuleti{\'c}}},\ }\href {\doibase
		10.1038/s41567-020-0917-6} {\bibfield  {journal} {\bibinfo  {journal} {Nature
				Physics}\ }\textbf {\bibinfo {volume} {16}},\ \bibinfo {pages} {921}
		(\bibinfo {year} {2020})}\BibitemShut {NoStop}%
	\bibitem [{\citenamefont {Thompson}\ \emph {et~al.}(2017)\citenamefont
		{Thompson}, \citenamefont {Nicholson}, \citenamefont {Liang}, \citenamefont
		{Cantu}, \citenamefont {Venkatramani}, \citenamefont {Choi}, \citenamefont
		{Fedorov}, \citenamefont {Viscor}, \citenamefont {Pohl}, \citenamefont
		{Lukin} \emph {et~al.}}]{Thompson:SymmetryProtectedCollisions:Nature2017}%
	\BibitemOpen
	\bibfield  {author} {\bibinfo {author} {\bibfnamefont {J.~D.}\ \bibnamefont
			{Thompson}}, \bibinfo {author} {\bibfnamefont {T.~L.}\ \bibnamefont
			{Nicholson}}, \bibinfo {author} {\bibfnamefont {Q.-Y.}\ \bibnamefont
			{Liang}}, \bibinfo {author} {\bibfnamefont {S.~H.}\ \bibnamefont {Cantu}},
		\bibinfo {author} {\bibfnamefont {A.~V.}\ \bibnamefont {Venkatramani}},
		\bibinfo {author} {\bibfnamefont {S.}~\bibnamefont {Choi}}, \bibinfo {author}
		{\bibfnamefont {I.~A.}\ \bibnamefont {Fedorov}}, \bibinfo {author}
		{\bibfnamefont {D.}~\bibnamefont {Viscor}}, \bibinfo {author} {\bibfnamefont
			{T.}~\bibnamefont {Pohl}}, \bibinfo {author} {\bibfnamefont {M.~D.}\
			\bibnamefont {Lukin}},  \emph {et~al.},\ }\href {\doibase
		10.1038/nature20823} {\bibfield  {journal} {\bibinfo  {journal} {Nature (London)}\
		}\textbf {\bibinfo {volume} {542}},\ \bibinfo {pages} {206} (\bibinfo {year}
		{2017})}\BibitemShut {NoStop}%
	\bibitem [{\citenamefont {Tiarks}\ \emph {et~al.}(2014)\citenamefont {Tiarks},
		\citenamefont {Baur}, \citenamefont {Schneider}, \citenamefont {D\"urr},\
		and\ \citenamefont {Rempe}}]{Tiarks:PhotonTransistor:PRL2014}%
	\BibitemOpen
	\bibfield  {author} {\bibinfo {author} {\bibfnamefont {D.}~\bibnamefont
			{Tiarks}}, \bibinfo {author} {\bibfnamefont {S.}~\bibnamefont {Baur}},
		\bibinfo {author} {\bibfnamefont {K.}~\bibnamefont {Schneider}}, \bibinfo
		{author} {\bibfnamefont {S.}~\bibnamefont {D\"urr}}, \ and\ \bibinfo {author}
		{\bibfnamefont {G.}~\bibnamefont {Rempe}},\ }\href {\doibase
		10.1103/PhysRevLett.113.053602} {\bibfield  {journal} {\bibinfo  {journal}
			{Phys. Rev. Lett.}\ }\textbf {\bibinfo {volume} {113}},\ \bibinfo {pages}
		{053602} (\bibinfo {year} {2014})}\BibitemShut {NoStop}%
	\bibitem [{\citenamefont {Gorniaczyk}\ \emph {et~al.}(2014)\citenamefont
		{Gorniaczyk}, \citenamefont {Tresp}, \citenamefont {Schmidt}, \citenamefont
		{Fedder},\ and\ \citenamefont
		{Hofferberth}}]{Gorniaczyk:PhotonTransistor:PRL2014}%
	\BibitemOpen
	\bibfield  {author} {\bibinfo {author} {\bibfnamefont {H.}~\bibnamefont
			{Gorniaczyk}}, \bibinfo {author} {\bibfnamefont {C.}~\bibnamefont {Tresp}},
		\bibinfo {author} {\bibfnamefont {J.}~\bibnamefont {Schmidt}}, \bibinfo
		{author} {\bibfnamefont {H.}~\bibnamefont {Fedder}}, \ and\ \bibinfo {author}
		{\bibfnamefont {S.}~\bibnamefont {Hofferberth}},\ }\href {\doibase
		10.1103/PhysRevLett.113.053601} {\bibfield  {journal} {\bibinfo  {journal}
			{Phys. Rev. Lett.}\ }\textbf {\bibinfo {volume} {113}},\ \bibinfo {pages}
		{053601} (\bibinfo {year} {2014})}\BibitemShut {NoStop}%
	\bibitem [{\citenamefont {Tiarks}\ \emph {et~al.}(2019)\citenamefont {Tiarks},
		\citenamefont {Schmidt-Eberle}, \citenamefont {Stolz}, \citenamefont
		{Rempe},\ and\ \citenamefont {D{\"u}rr}}]{Tiarks:QuantumGate:NatPhys19}%
	\BibitemOpen
	\bibfield  {author} {\bibinfo {author} {\bibfnamefont {D.}~\bibnamefont
			{Tiarks}}, \bibinfo {author} {\bibfnamefont {S.}~\bibnamefont
			{Schmidt-Eberle}}, \bibinfo {author} {\bibfnamefont {T.}~\bibnamefont
			{Stolz}}, \bibinfo {author} {\bibfnamefont {G.}~\bibnamefont {Rempe}}, \ and\
		\bibinfo {author} {\bibfnamefont {S.}~\bibnamefont {D{\"u}rr}},\ }\href
	{\doibase 10.1038/s41567-018-0313-7} {\bibfield  {journal} {\bibinfo
			{journal} {Nature Physics}\ }\textbf {\bibinfo {volume} {15}},\ \bibinfo
		{pages} {124} (\bibinfo {year} {2019})}\BibitemShut {NoStop}%
	\bibitem [{\citenamefont {Bienias}\ \emph {et~al.}(2020)\citenamefont
		{Bienias}, \citenamefont {Douglas}, \citenamefont {Paris-Mandoki},
		\citenamefont {Titum}, \citenamefont {Mirgorodskiy}, \citenamefont {Tresp},
		\citenamefont {Zeuthen}, \citenamefont {Gullans}, \citenamefont {Manzoni},
		\citenamefont {Hofferberth}, \citenamefont {Chang},\ and\ \citenamefont
		{Gorshkov}}]{Bienias:HighInput:PRR20}%
	\BibitemOpen
	\bibfield  {author} {\bibinfo {author} {\bibfnamefont {P.}~\bibnamefont
			{Bienias}}, \bibinfo {author} {\bibfnamefont {J.}~\bibnamefont {Douglas}},
		\bibinfo {author} {\bibfnamefont {A.}~\bibnamefont {Paris-Mandoki}}, \bibinfo
		{author} {\bibfnamefont {P.}~\bibnamefont {Titum}}, \bibinfo {author}
		{\bibfnamefont {I.}~\bibnamefont {Mirgorodskiy}}, \bibinfo {author}
		{\bibfnamefont {C.}~\bibnamefont {Tresp}}, \bibinfo {author} {\bibfnamefont
			{E.}~\bibnamefont {Zeuthen}}, \bibinfo {author} {\bibfnamefont {M.~J.}\
			\bibnamefont {Gullans}}, \bibinfo {author} {\bibfnamefont {M.}~\bibnamefont
			{Manzoni}}, \bibinfo {author} {\bibfnamefont {S.}~\bibnamefont
			{Hofferberth}}, \bibinfo {author} {\bibfnamefont {D.}~\bibnamefont {Chang}},
		\ and\ \bibinfo {author} {\bibfnamefont {A.~V.}\ \bibnamefont {Gorshkov}},\
	}\href {\doibase 10.1103/PhysRevResearch.2.033049} {\bibfield  {journal}
		{\bibinfo  {journal} {Phys. Rev. Research}\ }\textbf {\bibinfo {volume}
			{2}},\ \bibinfo {pages} {033049} (\bibinfo {year} {2020})}\BibitemShut
	{NoStop}%
	\bibitem [{\citenamefont {Pritchard}\ \emph {et~al.}(2010)\citenamefont
		{Pritchard}, \citenamefont {Maxwell}, \citenamefont {Gauguet}, \citenamefont
		{Weatherill}, \citenamefont {Jones},\ and\ \citenamefont
		{Adams}}]{Pritchard:firstEITmot:PRL10}%
	\BibitemOpen
	\bibfield  {author} {\bibinfo {author} {\bibfnamefont {J.~D.}\ \bibnamefont
			{Pritchard}}, \bibinfo {author} {\bibfnamefont {D.}~\bibnamefont {Maxwell}},
		\bibinfo {author} {\bibfnamefont {A.}~\bibnamefont {Gauguet}}, \bibinfo
		{author} {\bibfnamefont {K.~J.}\ \bibnamefont {Weatherill}}, \bibinfo
		{author} {\bibfnamefont {M.~P.~A.}\ \bibnamefont {Jones}}, \ and\ \bibinfo
		{author} {\bibfnamefont {C.~S.}\ \bibnamefont {Adams}},\ }\href {\doibase
		10.1103/PhysRevLett.105.193603} {\bibfield  {journal} {\bibinfo  {journal}
			{Phys. Rev. Lett.}\ }\textbf {\bibinfo {volume} {105}},\ \bibinfo {pages}
		{193603} (\bibinfo {year} {2010})}\BibitemShut {NoStop}%
	\bibitem [{\citenamefont {Schempp}\ \emph {et~al.}(2010)\citenamefont
		{Schempp}, \citenamefont {G\"unter}, \citenamefont {Hofmann}, \citenamefont
		{Giese}, \citenamefont {Saliba}, \citenamefont {DePaola}, \citenamefont
		{Amthor}, \citenamefont {Weidem\"uller}, \citenamefont
		{Sevin\ifmmode~\mbox{\c{c}}\else \c{c}\fi{}li},\ and\ \citenamefont
		{Pohl}}]{Schempp:CPT:PRL10}%
	\BibitemOpen
	\bibfield  {author} {\bibinfo {author} {\bibfnamefont {H.}~\bibnamefont
			{Schempp}}, \bibinfo {author} {\bibfnamefont {G.}~\bibnamefont {G\"unter}},
		\bibinfo {author} {\bibfnamefont {C.~S.}\ \bibnamefont {Hofmann}}, \bibinfo
		{author} {\bibfnamefont {C.}~\bibnamefont {Giese}}, \bibinfo {author}
		{\bibfnamefont {S.~D.}\ \bibnamefont {Saliba}}, \bibinfo {author}
		{\bibfnamefont {B.~D.}\ \bibnamefont {DePaola}}, \bibinfo {author}
		{\bibfnamefont {T.}~\bibnamefont {Amthor}}, \bibinfo {author} {\bibfnamefont
			{M.}~\bibnamefont {Weidem\"uller}}, \bibinfo {author} {\bibfnamefont
			{S.}~\bibnamefont {Sevin\ifmmode~\mbox{\c{c}}\else \c{c}\fi{}li}}, \ and\
		\bibinfo {author} {\bibfnamefont {T.}~\bibnamefont {Pohl}},\ }\href {\doibase
		10.1103/PhysRevLett.104.173602} {\bibfield  {journal} {\bibinfo  {journal}
			{Phys. Rev. Lett.}\ }\textbf {\bibinfo {volume} {104}},\ \bibinfo {pages}
		{173602} (\bibinfo {year} {2010})}\BibitemShut {NoStop}%
	\bibitem [{\citenamefont {Sevin\ifmmode~\mbox{\c{c}}\else \c{c}\fi{}li}\ \emph
		{et~al.}(2011{\natexlab{a}})\citenamefont {Sevin\ifmmode~\mbox{\c{c}}\else
			\c{c}\fi{}li}, \citenamefont {Ates}, \citenamefont {Pohl}, \citenamefont
		{Schempp}, \citenamefont {Hofmann}, \citenamefont {G{\"u}nter}, \citenamefont
		{Amthor}, \citenamefont {Weidem{\"u}ller}, \citenamefont {Pritchard},
		\citenamefont {Maxwell}, \citenamefont {Gauguet}, \citenamefont {Weatherill},
		\citenamefont {Jones},\ and\ \citenamefont
		{Adams}}]{Sevincli:Adams:CPTEIT:JoPB11}%
	\BibitemOpen
	\bibfield  {author} {\bibinfo {author} {\bibfnamefont {S.}~\bibnamefont
			{Sevin\ifmmode~\mbox{\c{c}}\else \c{c}\fi{}li}}, \bibinfo {author}
		{\bibfnamefont {C.}~\bibnamefont {Ates}}, \bibinfo {author} {\bibfnamefont
			{T.}~\bibnamefont {Pohl}}, \bibinfo {author} {\bibfnamefont {H.}~\bibnamefont
			{Schempp}}, \bibinfo {author} {\bibfnamefont {C.~S.}\ \bibnamefont
			{Hofmann}}, \bibinfo {author} {\bibfnamefont {G.}~\bibnamefont {G{\"u}nter}},
		\bibinfo {author} {\bibfnamefont {T.}~\bibnamefont {Amthor}}, \bibinfo
		{author} {\bibfnamefont {M.}~\bibnamefont {Weidem{\"u}ller}}, \bibinfo
		{author} {\bibfnamefont {J.~D.}\ \bibnamefont {Pritchard}}, \bibinfo {author}
		{\bibfnamefont {D.}~\bibnamefont {Maxwell}}, \bibinfo {author} {\bibfnamefont
			{A.}~\bibnamefont {Gauguet}}, \bibinfo {author} {\bibfnamefont {K.~J.}\
			\bibnamefont {Weatherill}}, \bibinfo {author} {\bibfnamefont {M.~P.~A.}\
			\bibnamefont {Jones}}, \ and\ \bibinfo {author} {\bibfnamefont {C.~S.}\
			\bibnamefont {Adams}},\ }\href
	{https://doi.org/10.1088/0953-4075/44/18/184018} {\bibfield  {journal}
		{\bibinfo  {journal} {J. Phys. B}\ }\textbf {\bibinfo {volume} {44}},\
		\bibinfo {pages} {184018} (\bibinfo {year} {2011}{\natexlab{a}})}\BibitemShut
	{NoStop}%
	\bibitem [{\citenamefont {Han}\ \emph {et~al.}(2016)\citenamefont {Han},
		\citenamefont {Vogt},\ and\ \citenamefont
		{Li}}]{Han:ShiftAndDephasing:PRA2016}%
	\BibitemOpen
	\bibfield  {author} {\bibinfo {author} {\bibfnamefont {J.}~\bibnamefont
			{Han}}, \bibinfo {author} {\bibfnamefont {T.}~\bibnamefont {Vogt}}, \ and\
		\bibinfo {author} {\bibfnamefont {W.}~\bibnamefont {Li}},\ }\href {\doibase
		10.1103/PhysRevA.94.043806} {\bibfield  {journal} {\bibinfo  {journal} {Phys.
				Rev. A}\ }\textbf {\bibinfo {volume} {94}},\ \bibinfo {pages} {043806}
		(\bibinfo {year} {2016})}\BibitemShut {NoStop}%
	\bibitem [{\citenamefont {DeSalvo}\ \emph {et~al.}(2016)\citenamefont
		{DeSalvo}, \citenamefont {Aman}, \citenamefont {Gaul}, \citenamefont {Pohl},
		\citenamefont {Yoshida}, \citenamefont {Burgd\"orfer}, \citenamefont
		{Hazzard}, \citenamefont {Dunning},\ and\ \citenamefont
		{Killian}}]{DeSalvo:MeanField:PRA2016}%
	\BibitemOpen
	\bibfield  {author} {\bibinfo {author} {\bibfnamefont {B.~J.}\ \bibnamefont
			{DeSalvo}}, \bibinfo {author} {\bibfnamefont {J.~A.}\ \bibnamefont {Aman}},
		\bibinfo {author} {\bibfnamefont {C.}~\bibnamefont {Gaul}}, \bibinfo {author}
		{\bibfnamefont {T.}~\bibnamefont {Pohl}}, \bibinfo {author} {\bibfnamefont
			{S.}~\bibnamefont {Yoshida}}, \bibinfo {author} {\bibfnamefont
			{J.}~\bibnamefont {Burgd\"orfer}}, \bibinfo {author} {\bibfnamefont
			{K.~R.~A.}\ \bibnamefont {Hazzard}}, \bibinfo {author} {\bibfnamefont
			{F.~B.}\ \bibnamefont {Dunning}}, \ and\ \bibinfo {author} {\bibfnamefont
			{T.~C.}\ \bibnamefont {Killian}},\ }\href {\doibase
		10.1103/PhysRevA.93.022709} {\bibfield  {journal} {\bibinfo  {journal} {Phys.
				Rev. A}\ }\textbf {\bibinfo {volume} {93}},\ \bibinfo {pages} {022709}
		(\bibinfo {year} {2016})}\BibitemShut {NoStop}%
	\bibitem [{\citenamefont {Sevin\ifmmode~\mbox{\c{c}}\else \c{c}\fi{}li}\ \emph
		{et~al.}(2011{\natexlab{b}})\citenamefont {Sevin\ifmmode~\mbox{\c{c}}\else
			\c{c}\fi{}li}, \citenamefont {Henkel}, \citenamefont {Ates},\ and\
		\citenamefont {Pohl}}]{Sevincli:adibatic:PRL11}%
	\BibitemOpen
	\bibfield  {author} {\bibinfo {author} {\bibfnamefont {S.}~\bibnamefont
			{Sevin\ifmmode~\mbox{\c{c}}\else \c{c}\fi{}li}}, \bibinfo {author}
		{\bibfnamefont {N.}~\bibnamefont {Henkel}}, \bibinfo {author} {\bibfnamefont
			{C.}~\bibnamefont {Ates}}, \ and\ \bibinfo {author} {\bibfnamefont
			{T.}~\bibnamefont {Pohl}},\ }\href {\doibase 10.1103/PhysRevLett.107.153001}
	{\bibfield  {journal} {\bibinfo  {journal} {Phys. Rev. Lett.}\ }\textbf
		{\bibinfo {volume} {107}},\ \bibinfo {pages} {153001} (\bibinfo {year}
		{2011}{\natexlab{b}})}\BibitemShut {NoStop}%
	\bibitem [{\citenamefont {Tebben}\ \emph {et~al.}(2019)\citenamefont {Tebben},
		\citenamefont {Hainaut}, \citenamefont {Walther}, \citenamefont {Zhang},
		\citenamefont {Z\"urn}, \citenamefont {Pohl},\ and\ \citenamefont
		{Weidem\"uller}}]{Tebben:ResonantEnhancement:PRA19}%
	\BibitemOpen
	\bibfield  {author} {\bibinfo {author} {\bibfnamefont {A.}~\bibnamefont
			{Tebben}}, \bibinfo {author} {\bibfnamefont {C.}~\bibnamefont {Hainaut}},
		\bibinfo {author} {\bibfnamefont {V.}~\bibnamefont {Walther}}, \bibinfo
		{author} {\bibfnamefont {Y.-C.}\ \bibnamefont {Zhang}}, \bibinfo {author}
		{\bibfnamefont {G.}~\bibnamefont {Z\"urn}}, \bibinfo {author} {\bibfnamefont
			{T.}~\bibnamefont {Pohl}}, \ and\ \bibinfo {author} {\bibfnamefont
			{M.}~\bibnamefont {Weidem\"uller}},\ }\href {\doibase
		10.1103/PhysRevA.100.063812} {\bibfield  {journal} {\bibinfo  {journal}
			{Phys. Rev. A}\ }\textbf {\bibinfo {volume} {100}},\ \bibinfo {pages}
		{063812} (\bibinfo {year} {2019})}\BibitemShut {NoStop}%
	\bibitem [{\citenamefont {G\"arttner}\ \emph
		{et~al.}(2014{\natexlab{a}})\citenamefont {G\"arttner}, \citenamefont
		{Whitlock}, \citenamefont {Sch\"onleber},\ and\ \citenamefont
		{Evers}}]{Gaerttner:Resonance:PRL14}%
	\BibitemOpen
	\bibfield  {author} {\bibinfo {author} {\bibfnamefont {M.}~\bibnamefont
			{G\"arttner}}, \bibinfo {author} {\bibfnamefont {S.}~\bibnamefont
			{Whitlock}}, \bibinfo {author} {\bibfnamefont {D.~W.}\ \bibnamefont
			{Sch\"onleber}}, \ and\ \bibinfo {author} {\bibfnamefont {J.}~\bibnamefont
			{Evers}},\ }\href {\doibase 10.1103/PhysRevLett.113.233002} {\bibfield
		{journal} {\bibinfo  {journal} {Phys. Rev. Lett.}\ }\textbf {\bibinfo
			{volume} {113}},\ \bibinfo {pages} {233002} (\bibinfo {year}
		{2014}{\natexlab{a}})}\BibitemShut {NoStop}%
	\bibitem [{\citenamefont {Gaul}\ \emph {et~al.}(2016)\citenamefont {Gaul},
		\citenamefont {DeSalvo}, \citenamefont {Aman}, \citenamefont {Dunning},
		\citenamefont {Killian},\ and\ \citenamefont
		{Pohl}}]{Gaul:ResDressing:PRL16}%
	\BibitemOpen
	\bibfield  {author} {\bibinfo {author} {\bibfnamefont {C.}~\bibnamefont
			{Gaul}}, \bibinfo {author} {\bibfnamefont {B.~J.}\ \bibnamefont {DeSalvo}},
		\bibinfo {author} {\bibfnamefont {J.~A.}\ \bibnamefont {Aman}}, \bibinfo
		{author} {\bibfnamefont {F.~B.}\ \bibnamefont {Dunning}}, \bibinfo {author}
		{\bibfnamefont {T.~C.}\ \bibnamefont {Killian}}, \ and\ \bibinfo {author}
		{\bibfnamefont {T.}~\bibnamefont {Pohl}},\ }\href {\doibase
		10.1103/PhysRevLett.116.243001} {\bibfield  {journal} {\bibinfo  {journal}
			{Phys. Rev. Lett.}\ }\textbf {\bibinfo {volume} {116}},\ \bibinfo {pages}
		{243001} (\bibinfo {year} {2016})}\BibitemShut {NoStop}%
	\bibitem [{\citenamefont {Helmrich}\ \emph {et~al.}(2016)\citenamefont
		{Helmrich}, \citenamefont {Arias}, \citenamefont {Pehoviak},\ and\
		\citenamefont {Whitlock}}]{Helmrich:ResDressing:PRL16}%
	\BibitemOpen
	\bibfield  {author} {\bibinfo {author} {\bibfnamefont {S.}~\bibnamefont
			{Helmrich}}, \bibinfo {author} {\bibfnamefont {A.}~\bibnamefont {Arias}},
		\bibinfo {author} {\bibfnamefont {N.}~\bibnamefont {Pehoviak}}, \ and\
		\bibinfo {author} {\bibfnamefont {S.}~\bibnamefont {Whitlock}},\ }\href
	{http://stacks.iop.org/0953-4075/49/i=3/a=03LT02} {\bibfield  {journal}
		{\bibinfo  {journal} {J. Phys. B}\ }\textbf {\bibinfo {volume} {49}},\
		\bibinfo {pages} {03LT02} (\bibinfo {year} {2016})}\BibitemShut {NoStop}%
	\bibitem [{\citenamefont {Weatherill}\ \emph {et~al.}(2008)\citenamefont
		{Weatherill}, \citenamefont {Pritchard}, \citenamefont {Abel}, \citenamefont
		{Bason}, \citenamefont {Mohapatra},\ and\ \citenamefont
		{Adams}}]{Weatherill_2008}%
	\BibitemOpen
	\bibfield  {author} {\bibinfo {author} {\bibfnamefont {K.~J.}\ \bibnamefont
			{Weatherill}}, \bibinfo {author} {\bibfnamefont {J.~D.}\ \bibnamefont
			{Pritchard}}, \bibinfo {author} {\bibfnamefont {R.~P.}\ \bibnamefont {Abel}},
		\bibinfo {author} {\bibfnamefont {M.~G.}\ \bibnamefont {Bason}}, \bibinfo
		{author} {\bibfnamefont {A.~K.}\ \bibnamefont {Mohapatra}}, \ and\ \bibinfo
		{author} {\bibfnamefont {C.~S.}\ \bibnamefont {Adams}},\ }\href {\doibase
		10.1088/0953-4075/41/20/201002} {\bibfield  {journal} {\bibinfo  {journal}
			{J. Phys. B}\ }\textbf
		{\bibinfo {volume} {41}},\ \bibinfo {pages} {201002} (\bibinfo {year}
		{2008})}\BibitemShut {NoStop}%
	\bibitem [{\citenamefont {Ates}\ \emph {et~al.}(2007)\citenamefont {Ates},
		\citenamefont {Pohl}, \citenamefont {Pattard},\ and\ \citenamefont
		{Rost}}]{Ates:ExcitationDynamicsRydberg:PRA07}%
	\BibitemOpen
	\bibfield  {author} {\bibinfo {author} {\bibfnamefont {C.}~\bibnamefont
			{Ates}}, \bibinfo {author} {\bibfnamefont {T.}~\bibnamefont {Pohl}}, \bibinfo
		{author} {\bibfnamefont {T.}~\bibnamefont {Pattard}}, \ and\ \bibinfo
		{author} {\bibfnamefont {J.~M.}\ \bibnamefont {Rost}},\ }\href {\doibase
		10.1103/PhysRevA.76.013413} {\bibfield  {journal} {\bibinfo  {journal} {Phys.
				Rev. A}\ }\textbf {\bibinfo {volume} {76}},\ \bibinfo {pages} {013413}
		(\bibinfo {year} {2007})}\BibitemShut {NoStop}%
	\bibitem [{\citenamefont {Ates}\ \emph {et~al.}(2011)\citenamefont {Ates},
		\citenamefont {Sevin\ifmmode~\mbox{\c{c}}\else \c{c}\fi{}li},\ and\
		\citenamefont {Pohl}}]{Ates:EITuniversality:PRA2011}%
	\BibitemOpen
	\bibfield  {author} {\bibinfo {author} {\bibfnamefont {C.}~\bibnamefont
			{Ates}}, \bibinfo {author} {\bibfnamefont {S.}~\bibnamefont
			{Sevin\ifmmode~\mbox{\c{c}}\else \c{c}\fi{}li}}, \ and\ \bibinfo {author}
		{\bibfnamefont {T.}~\bibnamefont {Pohl}},\ }\href {\doibase
		10.1103/PhysRevA.83.041802} {\bibfield  {journal} {\bibinfo  {journal} {Phys.
				Rev. A}\ }\textbf {\bibinfo {volume} {83}},\ \bibinfo {pages} {041802(R)}
		(\bibinfo {year} {2011})}\BibitemShut {NoStop}%
	\bibitem [{\citenamefont {Heeg}\ \emph {et~al.}(2012)\citenamefont {Heeg},
		\citenamefont {G\"arttner},\ and\ \citenamefont
		{Evers}}]{Heeg:Evers:HybridModel:PRA12}%
	\BibitemOpen
	\bibfield  {author} {\bibinfo {author} {\bibfnamefont {K.~P.}\ \bibnamefont
			{Heeg}}, \bibinfo {author} {\bibfnamefont {M.}~\bibnamefont {G\"arttner}}, \
		and\ \bibinfo {author} {\bibfnamefont {J.}~\bibnamefont {Evers}},\ }\href
	{\doibase 10.1103/PhysRevA.86.063421} {\bibfield  {journal} {\bibinfo
			{journal} {Phys. Rev. A}\ }\textbf {\bibinfo {volume} {86}},\ \bibinfo
		{pages} {063421} (\bibinfo {year} {2012})}\BibitemShut {NoStop}%
	\bibitem [{\citenamefont {G\"arttner}\ and\ \citenamefont
		{Evers}(2013)}]{Gaerttner:REwithPropagation:PRA13}%
	\BibitemOpen
	\bibfield  {author} {\bibinfo {author} {\bibfnamefont {M.}~\bibnamefont
			{G\"arttner}}\ and\ \bibinfo {author} {\bibfnamefont {J.}~\bibnamefont
			{Evers}},\ }\href {\doibase 10.1103/PhysRevA.88.033417} {\bibfield  {journal}
		{\bibinfo  {journal} {Phys. Rev. A}\ }\textbf {\bibinfo {volume} {88}},\
		\bibinfo {pages} {033417} (\bibinfo {year} {2013})}\BibitemShut {NoStop}%
	\bibitem [{\citenamefont {Han}\ \emph {et~al.}(2015)\citenamefont {Han},
		\citenamefont {Vogt}, \citenamefont {Manjappa}, \citenamefont {Guo},
		\citenamefont {Kiffner},\ and\ \citenamefont {Li}}]{Li:EITlensing:PRA15}%
	\BibitemOpen
	\bibfield  {author} {\bibinfo {author} {\bibfnamefont {J.}~\bibnamefont
			{Han}}, \bibinfo {author} {\bibfnamefont {T.}~\bibnamefont {Vogt}}, \bibinfo
		{author} {\bibfnamefont {M.}~\bibnamefont {Manjappa}}, \bibinfo {author}
		{\bibfnamefont {R.}~\bibnamefont {Guo}}, \bibinfo {author} {\bibfnamefont
			{M.}~\bibnamefont {Kiffner}}, \ and\ \bibinfo {author} {\bibfnamefont
			{W.}~\bibnamefont {Li}},\ }\href {\doibase 10.1103/PhysRevA.92.063824}
	{\bibfield  {journal} {\bibinfo  {journal} {Phys. Rev. A}\ }\textbf {\bibinfo
			{volume} {92}},\ \bibinfo {pages} {063824} (\bibinfo {year}
		{2015})}\BibitemShut {NoStop}%
	\bibitem [{\citenamefont {Robert-de Saint-Vincent}\ \emph
		{et~al.}(2013)\citenamefont {Robert-de Saint-Vincent}, \citenamefont
		{Hofmann}, \citenamefont {Schempp}, \citenamefont {G\"unter}, \citenamefont
		{Whitlock},\ and\ \citenamefont
		{Weidem\"uller}}]{RobertVincent:AvalancheIonization:PRL13}%
	\BibitemOpen
	\bibfield  {author} {\bibinfo {author} {\bibfnamefont {M.}~\bibnamefont
			{Robert-de-Saint-Vincent}}, \bibinfo {author} {\bibfnamefont {C.~S.}\
			\bibnamefont {Hofmann}}, \bibinfo {author} {\bibfnamefont {H.}~\bibnamefont
			{Schempp}}, \bibinfo {author} {\bibfnamefont {G.}~\bibnamefont {G\"unter}},
		\bibinfo {author} {\bibfnamefont {S.}~\bibnamefont {Whitlock}}, \ and\
		\bibinfo {author} {\bibfnamefont {M.}~\bibnamefont {Weidem\"uller}},\ }\href
	{\doibase 10.1103/PhysRevLett.110.045004} {\bibfield  {journal} {\bibinfo
			{journal} {Phys. Rev. Lett.}\ }\textbf {\bibinfo {volume} {110}},\ \bibinfo
		{pages} {045004} (\bibinfo {year} {2013})}\BibitemShut {NoStop}%
	\bibitem [{\citenamefont {Fleischhauer}\ \emph {et~al.}(2005)\citenamefont
		{Fleischhauer}, \citenamefont {Imamoglu},\ and\ \citenamefont
		{Marangos}}]{Fleischhauer:EIT:PMP05}%
	\BibitemOpen
	\bibfield  {author} {\bibinfo {author} {\bibfnamefont {M.}~\bibnamefont
			{Fleischhauer}}, \bibinfo {author} {\bibfnamefont {A.}~\bibnamefont
			{Imamoglu}}, \ and\ \bibinfo {author} {\bibfnamefont {J.~P.}\ \bibnamefont
			{Marangos}},\ }\href {\doibase 10.1103/RevModPhys.77.633} {\bibfield
		{journal} {\bibinfo  {journal} {Rev. Mod. Phys.}\ }\textbf {\bibinfo {volume}
			{77}},\ \bibinfo {pages} {633} (\bibinfo {year} {2005})}\BibitemShut
	{NoStop}%
	\bibitem [{\citenamefont {Hsiao}\ \emph {et~al.}(2020)\citenamefont {Hsiao},
		\citenamefont {Chen},\ and\ \citenamefont {Yu}}]{Hsiao:MFnearestNeighbor:20}%
	\BibitemOpen
	\bibfield  {author} {\bibinfo {author} {\bibfnamefont {S.-S.}\ \bibnamefont
			{Hsiao}}, \bibinfo {author} {\bibfnamefont {K.-T.}\ \bibnamefont {Chen}}, \
		and\ \bibinfo {author} {\bibfnamefont {I.~A.}\ \bibnamefont {Yu}},\ }\href
	{\doibase 10.1364/OE.401310} {\bibfield  {journal} {\bibinfo  {journal} {Opt.
				Express}\ }\textbf {\bibinfo {volume} {28}},\ \bibinfo {pages} {28414}
		(\bibinfo {year} {2020})}\BibitemShut {NoStop}%
	\bibitem [{\citenamefont {G\"arttner}\ \emph
		{et~al.}(2014{\natexlab{b}})\citenamefont {G\"arttner}, \citenamefont
		{Whitlock}, \citenamefont {Sch\"onleber},\ and\ \citenamefont
		{Evers}}]{Gaerttner:SemianalyticalModel:PRA14}%
	\BibitemOpen
	\bibfield  {author} {\bibinfo {author} {\bibfnamefont {M.}~\bibnamefont
			{G\"arttner}}, \bibinfo {author} {\bibfnamefont {S.}~\bibnamefont
			{Whitlock}}, \bibinfo {author} {\bibfnamefont {D.~W.}\ \bibnamefont
			{Sch\"onleber}}, \ and\ \bibinfo {author} {\bibfnamefont {J.}~\bibnamefont
			{Evers}},\ }\href {\doibase 10.1103/PhysRevA.89.063407} {\bibfield  {journal}
		{\bibinfo  {journal} {Phys. Rev. A}\ }\textbf {\bibinfo {volume} {89}},\
		\bibinfo {pages} {063407} (\bibinfo {year} {2014}{\natexlab{b}})}\BibitemShut
	{NoStop}%
	\bibitem [{\citenamefont {Hofmann}\ \emph {et~al.}(2014)\citenamefont
		{Hofmann}, \citenamefont {G{\"u}nter}, \citenamefont {Schempp}, \citenamefont
		{M{\"u}ller}, \citenamefont {Faber}, \citenamefont {Busche}, \citenamefont
		{Robert-de Saint-Vincent}, \citenamefont {Whitlock},\ and\ \citenamefont
		{Weidem{\"u}ller}}]{Hofmann:setup:Frontiers14}%
	\BibitemOpen
	\bibfield  {author} {\bibinfo {author} {\bibfnamefont {C.~S.}\ \bibnamefont
			{Hofmann}}, \bibinfo {author} {\bibfnamefont {G.}~\bibnamefont {G{\"u}nter}},
		\bibinfo {author} {\bibfnamefont {H.}~\bibnamefont {Schempp}}, \bibinfo
		{author} {\bibfnamefont {N.~L.~M.}\ \bibnamefont {M{\"u}ller}}, \bibinfo
		{author} {\bibfnamefont {A.}~\bibnamefont {Faber}}, \bibinfo {author}
		{\bibfnamefont {H.}~\bibnamefont {Busche}}, \bibinfo {author} {\bibfnamefont
			{M.}~\bibnamefont {Robert-de Saint-Vincent}}, \bibinfo {author}
		{\bibfnamefont {S.}~\bibnamefont {Whitlock}}, \ and\ \bibinfo {author}
		{\bibfnamefont {M.}~\bibnamefont {Weidem{\"u}ller}},\ }\href {\doibase
		10.1007/s11467-013-0396-7} {\bibfield  {journal} {\bibinfo  {journal}
			{Frontiers of Physics}\ }\textbf {\bibinfo {volume} {9}},\ \bibinfo {pages}
		{571} (\bibinfo {year} {2014})}\BibitemShut {NoStop}%
	\bibitem [{\citenamefont {Petrich}\ \emph {et~al.}(1994)\citenamefont
		{Petrich}, \citenamefont {Anderson}, \citenamefont {Ensher},\ and\
		\citenamefont {Cornell}}]{Petrich:compressedMOT:JosaB94}%
	\BibitemOpen
	\bibfield  {author} {\bibinfo {author} {\bibfnamefont {W.}~\bibnamefont
			{Petrich}}, \bibinfo {author} {\bibfnamefont {M.~H.}\ \bibnamefont
			{Anderson}}, \bibinfo {author} {\bibfnamefont {J.~R.}\ \bibnamefont
			{Ensher}}, \ and\ \bibinfo {author} {\bibfnamefont {E.~A.}\ \bibnamefont
			{Cornell}},\ }\href {\doibase 10.1364/JOSAB.11.001332} {\bibfield  {journal}
		{\bibinfo  {journal} {J. Opt. Soc. Am. B}\ }\textbf {\bibinfo {volume}
			{11}},\ \bibinfo {pages} {1332} (\bibinfo {year} {1994})}\BibitemShut
	{NoStop}%
	\bibitem [{\citenamefont {Townsend}\ \emph {et~al.}(1996)\citenamefont
		{Townsend}, \citenamefont {Edwards}, \citenamefont {Zetie}, \citenamefont
		{Cooper}, \citenamefont {Rink},\ and\ \citenamefont
		{Foot}}]{Townsend:darkMOT:PRA96}%
	\BibitemOpen
	\bibfield  {author} {\bibinfo {author} {\bibfnamefont {C.~G.}\ \bibnamefont
			{Townsend}}, \bibinfo {author} {\bibfnamefont {N.~H.}\ \bibnamefont
			{Edwards}}, \bibinfo {author} {\bibfnamefont {K.~P.}\ \bibnamefont {Zetie}},
		\bibinfo {author} {\bibfnamefont {C.~J.}\ \bibnamefont {Cooper}}, \bibinfo
		{author} {\bibfnamefont {J.}~\bibnamefont {Rink}}, \ and\ \bibinfo {author}
		{\bibfnamefont {C.~J.}\ \bibnamefont {Foot}},\ }\href {\doibase
		10.1103/PhysRevA.53.1702} {\bibfield  {journal} {\bibinfo  {journal} {Phys.
				Rev. A}\ }\textbf {\bibinfo {volume} {53}},\ \bibinfo {pages} {1702}
		(\bibinfo {year} {1996})}\BibitemShut {NoStop}%
	\bibitem [{\citenamefont {Hertel}\ and\ \citenamefont
		{Schulz}(2015)}]{Hertel:AMOPhysics:Springer15}%
	\BibitemOpen
	\bibfield  {author} {\bibinfo {author} {\bibfnamefont {I.~V.}\ \bibnamefont
			{Hertel}}\ and\ \bibinfo {author} {\bibfnamefont {C.-P.}\ \bibnamefont
			{Schulz}},\ }\href {\doibase 10.1007/978-3-642-54322-7} {\emph {\bibinfo
			{title} {Atoms, Molecules and Optical Physics 1: Atoms and
				Spectroscopy}}}\ (\bibinfo  {publisher}
	{Springer, Berlin},\ \bibinfo {year} {2015})\BibitemShut {NoStop}%
	\bibitem [{\citenamefont {Labeyrie}\ \emph {et~al.}(2003)\citenamefont
		{Labeyrie}, \citenamefont {Vaujour}, \citenamefont {M\"uller}, \citenamefont
		{Delande}, \citenamefont {Miniatura}, \citenamefont {Wilkowski},\ and\
		\citenamefont {Kaiser}}]{Labeyrie:SlowDiffusionLight:PRL03}%
	\BibitemOpen
	\bibfield  {author} {\bibinfo {author} {\bibfnamefont {G.}~\bibnamefont
			{Labeyrie}}, \bibinfo {author} {\bibfnamefont {E.}~\bibnamefont {Vaujour}},
		\bibinfo {author} {\bibfnamefont {C.~A.}\ \bibnamefont {M\"uller}}, \bibinfo
		{author} {\bibfnamefont {D.}~\bibnamefont {Delande}}, \bibinfo {author}
		{\bibfnamefont {C.}~\bibnamefont {Miniatura}}, \bibinfo {author}
		{\bibfnamefont {D.}~\bibnamefont {Wilkowski}}, \ and\ \bibinfo {author}
		{\bibfnamefont {R.}~\bibnamefont {Kaiser}},\ }\href {\doibase
		10.1103/PhysRevLett.91.223904} {\bibfield  {journal} {\bibinfo  {journal}
			{Phys. Rev. Lett.}\ }\textbf {\bibinfo {volume} {91}},\ \bibinfo {pages}
		{223904} (\bibinfo {year} {2003})}\BibitemShut {NoStop}%
	\bibitem [{\citenamefont {Labeyrie}\ \emph {et~al.}(2005)\citenamefont
		{Labeyrie}, \citenamefont {Kaiser},\ and\ \citenamefont
		{Delande}}]{Labeyrie:RadiationTrapping:APB05}%
	\BibitemOpen
	\bibfield  {author} {\bibinfo {author} {\bibfnamefont {G.}~\bibnamefont
			{Labeyrie}}, \bibinfo {author} {\bibfnamefont {R.}~\bibnamefont {Kaiser}}, \
		and\ \bibinfo {author} {\bibfnamefont {D.}~\bibnamefont {Delande}},\ }\href
	{\doibase 10.1007/s00340-005-2015-y} {\bibfield  {journal} {\bibinfo
			{journal} {Applied Physics B}\ }\textbf {\bibinfo {volume} {81}},\ \bibinfo
		{pages} {1001} (\bibinfo {year} {2005})}\BibitemShut {NoStop}%
	\bibitem [{\citenamefont {Ferreira-Cao}\ \emph {et~al.}(2020)\citenamefont
		{Ferreira-Cao}, \citenamefont {Gavryusev}, \citenamefont {Franz},
		\citenamefont {Alves}, \citenamefont {Signoles}, \citenamefont {Z\"urn},\
		and\ \citenamefont {Weidem\"uller}}]{Ferreira:DepletionImaging:JPhysB20}%
	\BibitemOpen
	\bibfield  {author} {\bibinfo {author} {\bibfnamefont {M.}~\bibnamefont
			{Ferreira-Cao}}, \bibinfo {author} {\bibfnamefont {V.}~\bibnamefont
			{Gavryusev}}, \bibinfo {author} {\bibfnamefont {T.}~\bibnamefont {Franz}},
		\bibinfo {author} {\bibfnamefont {R.~F.}\ \bibnamefont {Alves}}, \bibinfo
		{author} {\bibfnamefont {A.}~\bibnamefont {Signoles}}, \bibinfo {author}
		{\bibfnamefont {G.}~\bibnamefont {Z\"urn}}, \ and\ \bibinfo {author}
		{\bibfnamefont {M.}~\bibnamefont {Weidem\"uller}},\ }\href {\doibase
		10.1088/1361-6455/ab7427} {\bibfield  {journal} {\bibinfo  {journal} {J.
				Phys. B}\ }\textbf {\bibinfo {volume} {53}},\ \bibinfo {pages} {084004}
		(\bibinfo {year} {2020})}\BibitemShut {NoStop}%
	\bibitem [{\citenamefont {Sadler}\ \emph {et~al.}(2017)\citenamefont {Sadler},
		\citenamefont {Bridge}, \citenamefont {Boddy}, \citenamefont {Bounds},
		\citenamefont {Keegan}, \citenamefont {Lochead}, \citenamefont {Jones},\ and\
		\citenamefont {Olmos}}]{Sadler:RadiationTrappingPollutans:PRA17}%
	\BibitemOpen
	\bibfield  {author} {\bibinfo {author} {\bibfnamefont {D.~P.}\ \bibnamefont
			{Sadler}}, \bibinfo {author} {\bibfnamefont {E.~M.}\ \bibnamefont {Bridge}},
		\bibinfo {author} {\bibfnamefont {D.}~\bibnamefont {Boddy}}, \bibinfo
		{author} {\bibfnamefont {A.~D.}\ \bibnamefont {Bounds}}, \bibinfo {author}
		{\bibfnamefont {N.~C.}\ \bibnamefont {Keegan}}, \bibinfo {author}
		{\bibfnamefont {G.}~\bibnamefont {Lochead}}, \bibinfo {author} {\bibfnamefont
			{M.~P.~A.}\ \bibnamefont {Jones}}, \ and\ \bibinfo {author} {\bibfnamefont
			{B.}~\bibnamefont {Olmos}},\ }\href {\doibase 10.1103/PhysRevA.95.013839}
	{\bibfield  {journal} {\bibinfo  {journal} {Phys. Rev. A}\ }\textbf {\bibinfo
			{volume} {95}},\ \bibinfo {pages} {013839} (\bibinfo {year}
		{2017})}\BibitemShut {NoStop}%
\end{thebibliography}
\end{document}